\documentclass[11pt, a4paper]{elsarticle}

\usepackage{pdfsync}
\usepackage{a4wide}
\usepackage{graphicx}
\usepackage{graphics}
\usepackage{amssymb,latexsym}
\usepackage[scaled=0.95]{helvet}
\usepackage{amsfonts}
\usepackage{amsmath}
\usepackage{multicol}
\usepackage{textcomp}
\usepackage{indentfirst}
\usepackage{array}
\usepackage{nccmath}
\usepackage[T1]{fontenc}
\usepackage[utf8]{inputenc}
\usepackage{verbatim}
\usepackage{courier}
\usepackage[english]{babel}
\usepackage{theorem}

\usepackage{tikz}
\usepackage{epsfig,psfrag}

\usepackage{natbib}
\setcitestyle{authoryear,open={(},close={)}, square, comma, sort&compress}
\usepackage[breaklinks=true, colorlinks=false, pdfborder={0 0 0}, bookmarks=false]{hyperref}
\usepackage{ragged2e}
\usepackage{url}

\newtheorem{theorem}{Theorem}

\newtheorem{proposition}[theorem]{Proposition}
\newtheorem{defi}[theorem]{Definition}

\newtheorem{remark}[theorem]{Remark}
\newtheorem{cor}[theorem]{Corollary}
\allowdisplaybreaks

\begin{document}

\begin{frontmatter}

\journal{Theoretical Population Biology}
\title{{Looking down in the ancestral selection graph:\\
 A probabilistic approach to the common ancestor type distribution}}

\author[label1]{Ute Lenz\fnref{fn1}}
\author[label2]{Sandra Kluth\fnref{fn2}}
\author[label2]{Ellen Baake\corref{cor1}}
\author[label1]{Anton Wakolbinger\fnref{fn3}}
\address[label1]{Institut f\"ur Mathematik, Goethe-Universit\"at Frankfurt, Box 111932, 60054 Frankfurt am Main, Germany}
\address[label2]{Faculty of Technology, Bielefeld University, Box 100131, 33501 Bielefeld, Germany}

\cortext[cor1]{Corresponding author. Phone: +49--521--106--4896. E-mail: ebaake@techfak.uni-bielefeld.de (E.~Baake). }
\fntext[fn1]{E-mail: lenz@math.uni-frankfurt.de.}
\fntext[fn2]{Present address: Department of Internal Medicine I and Center for Integrated Oncology, University Hospital Cologne, Kerpener Strasse 62, 50937 Koeln. E-mail: sandra.kluth@uni-bielefeld.de.}
\fntext[fn3]{E-mail: wakolbin@math.uni-frankfurt.de.}

\begin{abstract}
In a  (two-type) Wright-Fisher diffusion with directional selection and two-way mutation,  
let $x$ denote today's frequency of the beneficial type, and given $x$, let $h(x)$ be the
probability that, among all individuals of today's population, the individual whose progeny
will eventually take over in the population is of the beneficial type. Fearnhead 
[Fearnhead, P., 2002. The common ancestor at a nonneutral locus. J. Appl. Probab. 39, 38-54]
and Taylor
[Taylor, J. E., 2007. The common ancestor process for a Wright-Fisher diffusion. Electron. J. Probab. 12, 808-847] obtained a series representation for $h(x)$. We develop a  
construction that contains elements of both the ancestral selection graph
and the lookdown construction and  includes pruning of certain lines
upon mutation.
Besides being interesting in its
own right, this construction allows a transparent derivation of the 
series coefficients of $h(x)$ and gives them a probabilistic meaning.

\end{abstract}


\end{frontmatter}

\section{Introduction}
The understanding of ancestral processes under selection and mutation is among the
fundamental challenges in population genetics. 
Two central concepts  are the 
ancestral selection graph (ASG) and the
lookdown (LD) construction. 
The ancestral selection graph  (\citealt*{Krone}; 
\citealt*{Neuhauser}; see also 
\citealt*{Shiga} for an analogous construction in a diffusion model with spatial structure) describes the set of
lines that are potential ancestors of  a sample of individuals
taken from a present population. In contrast, 
the lookdown construction
\citep{DonnellyKurtz99,DonnellyKurtz99b} is 
an integrated  representation that
makes all individual lines in a population 
explicit, together with the genealogies of arbitrary samples.  
See \citet*[Chapter 5]{Etheridge} for an excellent overview of the area.

Both  the ASG and the LD are important   theoretical
concepts as well as  
valuable
 tools in applications.
Interest is usually directed towards the genealogy of a sample, 
backwards in time  until the most recent common ancestor (MRCA).
However, the ancestral line that continues beyond the MRCA into the
distant past is of considerable interest on its own, not least 
because it links the genealogy (of a sample from a  population)
to the longer time scale of phylogenetic trees. 
The extended time horizon then
shifts attention to the asymptotic properties of the ancestral process. The stationary type distribution on the ancestral line
may differ substantially from the  stationary type distribution in the population.
This mirrors the fact that the ancestral
line consists of those
individuals that are successful in the long run; thus, its type distribution 
is expected to be biased towards the favourable types.

When looking at the evolution of the system in (forward) time $[0,\infty)$, one may ask for properties of the so-called {\em immortal line}, which is the line of descent of those  individuals whose offspring eventually takes over the entire population. In other words, the immortal line restricted to any time interval $[0,t]$ is the common ancestral line of the population back from the far future. It then makes sense to consider the type of the immortal line at time $0$. To be specific, let us consider a Wright-Fisher diffusion with two types of which one is more and one is less fit. The {\em common ancestor type (CAT) distribution} at time $0$, conditional on the type frequencies $(x,1-x)$, then has weights $(h(x),1-h(x))$, where $h(x)$ is the probability that  the population ultimately consists
of offspring of  an individual of the beneficial type, when starting with a frequency $x$ of beneficial individuals at time $0$. 

The quantity $h(x)$ can also be understood as the limiting probability (as $s\to \infty$) that the ancestor at time $0$ of an individual sampled from the population at the future time $s$ is of the beneficial type, given that the frequency of the beneficial type at time $0$ is $x$. Equivalently, $h(x)$ is the limiting probability (as $s\to \infty$) that the ancestor at the past time $-s$ of an individual sampled from the population at  time $0$ is of the beneficial type, given that the frequency of the beneficial type at time $-s$ was $x$.

\citet*{Fearnhead} computed the common ancestor type distribution for time-stationary type frequencies, representing it in the form $\int_0^1 (h(x), 1-h(x)) \pi(dx)$ (where $\pi$ is  Wright's equilibrium distribution) and calculating a recursion for the coefficients of a series representation of $h(x)$. Later, $h(x)$ has been represented in terms of a boundary value problem \citep{Taylor,Kluth}, see also Section~\ref{sec:bvp}.

In the case without mutations (in which $h(x)$  coincides with the classical fixation probability  of the beneficial type starting from frequency $x$), \citet{Mano} and \citet*{Pokalyuk} have represented
 $h(x)$ in terms of the {\em equilibrium ASG}, making use of a time reversal argument (see Section \ref{secASG}). 
 However, the generalisation to the case with mutation
is anything but obvious. One purpose of this article is to solve this problem.
A key ingredient will be a combination of the
ASG with elements of the lookdown construction, which also seems of interest in its own right.

  The paper is organised as follows.
In Section~2, we start by briefly recapitulating the ASG (starting from the Moran model
for definiteness). We then recall the Fearnhead-Taylor representation of $h(x)$ and give its explanation in terms of the equilibrium ASG in the case without mutations, inspired by \citet*{Pokalyuk}. In Section~3, we prepare the scene by ordering the lines of the ASG
in a specific way; in Section~4, we then represent the ordered ASG in terms of a
fixed arrangement of levels, akin to a lookdown construction. In Section~5, a pruning 
procedure is described that reduces the number of lines upon mutation. 
The stationary number of lines in the resulting pruned LD-ASG will provide the
desired connection to the (conditional) common ancestor type distribution. 
Namely, the tail probabilities of the number of lines appear as the
coefficients in the series representation. In Section~6, the graphical approach will directly reveal
various monotonicity properties of the tail probabilities as functions of the model parameters,
which translate into monotonicity properties of the common ancestor type distribution. Section~7 is an add-on, which
makes the connection to Taylor's boundary value problem for $h(x)$ explicit;
Section~8 contains some concluding remarks.

\section{Concepts and models}

\subsection{The Moran model and its diffusion limit}
\label{moranmodelsection}

Let us consider a haploid population of fixed size $N \in \mathbb{N}$ in which each individual is characterised
by a type $i \in S:= \{ 0,1 \}$. An individual of type $i$ may, at any instant in continuous time, 
do either of two things: it may reproduce, which happens at rate $1$ if $i=1$ and
at rate $1+s_N^{}$, $s_N^{} \geq 0$, if $i=0$; or it may mutate to type $j$ at rate
$u_N^{} \nu_j^{}$, $u_N^{} \geq 0$, $0 \leq \nu_j^{} \leq 1$, $\nu_0^{}
+\nu_1^{} =1$. 
If an individual reproduces, its single offspring inherits the 
parent's type and replaces a randomly chosen individual, maybe its own parent.
Concerning mutations, $u_N^{}$ is the total mutation rate and $\nu_j^{}$ the probability
of a mutation to type $j$. Note that the possibility of silent mutations from type $j$ to type $j$ is included.

\begin{figure}[ht]
\begin{center}
\psfrag{t}{\LARGE$t$}
\psfrag{0}{\Large$0$}
\psfrag{1}{\Large$1$}
\includegraphics[angle=0, width=0.6\textwidth]{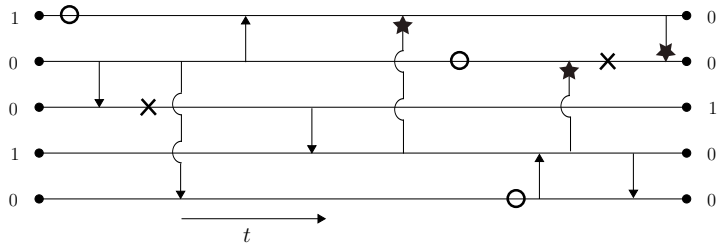}
\caption{The Moran model with two-way mutation and selection. The types are indicated for the initial population (left) and the final one (right). Crosses represent mutations to type $1$, circles mutations to type $0$. Selective events are depicted as arrows with star-shaped heads.}
\label{MoranModel}
\end{center}
\end{figure}
The Moran model has a well-known graphical illustration as an interacting particle system (cf. Fig.~\ref{MoranModel}).
The individuals are represented by horizontal line pieces, with forward time running from left to right in the figure.
 Arrows indicate
reproduction events with the parent at its tail and the offspring at its head. 
For later use, we decompose reproduction events into neutral and selective ones. Neutral arrows appear 
at rate $1/N$, selective arrows (those with a star-shaped arrowhead in Fig.~\ref{MoranModel}) at rate $s_N^{}/N$ per ordered pair of lines, 
irrespective of their types. The rates specified above are obtained by the convention that neutral arrows
may be used by all individuals, whereas  selective  arrows may only be used by type-$0$ individuals and are ignored otherwise.
Mutations to type $0$ are marked by circles, mutations to type $1$ by crosses.

The usual diffusion rescaling in population genetics is applied, i.e. rates are rescaled such that 
$\lim_{N \to \infty} N s_N^{} = \sigma$ and 
$\lim_{N \to \infty} N u_N^{} = \theta$, $0 \leq \sigma, \theta < \infty$, and time is sped up by a 
factor of $N$. Let $X_t^{}$ be the 
frequency of type-$0$ individuals at time $t$ in this diffusion limit. Then, the process $(X_t^{})_{t \in \mathbb{R}}$ is a Wright-Fisher diffusion which is characterised by
the drift coefficient
$a(x) = (1-x) \theta \nu_0^{} - x \theta \nu_1^{} + x(1-x) \sigma$
and the diffusion coefficient
$b(x) = 2x(1-x).$
The stationary density $\pi$ is given by
$ \pi(x) = C (1-x)^{\theta \nu_1 -1} x^{\theta \nu_0 -1} \exp (\sigma x), $
where $C$ is a normalising constant (cf. \citet[Chapters 7, 8]{Durrett} or \citet[Chapters 4, 5]{Ewens}).

\subsection{The ancestral selection graph}\label{secASG}
\begin{figure}[ht]
\begin{center}
\psfrag{I}{\Large$I$}
\psfrag{C}{\Large$C$}
\psfrag{D}{\Large$D$}
\psfrag{0}{\Large$0$}
\psfrag{1}{\Large$1$}
\includegraphics[angle=0,width=8.8cm, height=3.5cm]{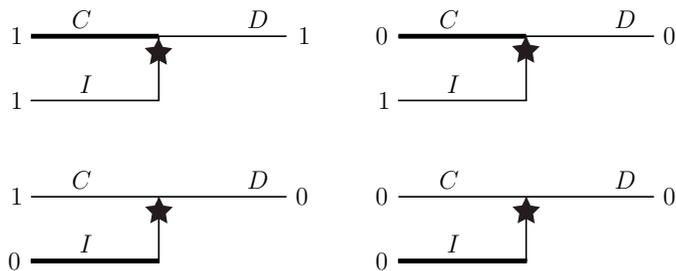}
\caption{Incoming branch ($I$), continuing branch ($C$), and descendant ($D$). The ancestor is marked bold.}
\label{IC}
\end{center}
\end{figure}

The \textit{ancestral selection graph} was introduced by \citet*{Krone} and \citet*{Neuhauser} to construct
samples from a present population, together with their ancestries,
in the diffusion limit of the Moran model with mutation and selection. The basic idea is to 
understand selective arrows 
as unresolved reproduction events backwards in time: 
the descendant has two \textit{potential ancestors}, the 
\textit{incoming branch} (at the tail) and the
\textit{continuing branch} (at the tip), see also Fig.~\ref{IC}. 
The incoming 
branch is the ancestor if it is of type $0$, otherwise the continuing one is ancestral. For a hands-on exposition, see \citet[Chapter 7.1]{Wakeley}.

The ASG is constructed by starting from the (as
yet untyped) sample and tracing back the lines of all potential ancestors. In the finite graphical representation,
a neutral arrow that joins two potential ancestral  lines appears at rate 
$2/N$ per currently extant pair of potential ancestral lines, 
then giving rise to a \textit{coalescence event}, i.e.
the two lines merge into a single one. In the same finite setting, a selective arrow that
emanates from outside the current set of $n$ potential ancestral lines and
hits this set
appears at rate $n (N-n) s_N^{} /N$. This gives rise to a \textit{branching event}, i.e., viewed backwards in time, the line that is hit by the selective arrow splits into an
incoming and continuing branch as described above. Thus, in the diffusion limit, 
since $N-n \sim N$ as $N\to \infty$, 
the process $(K_{r}^{})_{r \in \mathbb{R}}^{}$, where $K_{r}$ is the number of lines in the ASG at time $r=-t$, evolves backwards in time with rates\footnote{Since our population size is $N$ (rather than $2N$), the selection 
coefficient $\sigma$ in our scenario corresponds to $\sigma / 2$ and our $n(n-1)$ to 
$n(n-1) / 2$ in Krone and Neuhauser (1997).}
\begin{equation}
{q}_K(n,n-1) = n(n-1) , \quad
{q}_K(n,n+1) = n\sigma,  \qquad n\in \mathbb N.
\label{ratesASGNKbackwards}
\end{equation}
At a coalescence event a randomly chosen pair of lines  coalesces, while at a branching event
a randomly chosen line splits into two.

The (reversible) equilibrium distribution of the  dynamics \eqref{ratesASGNKbackwards} turns out to be the Poisson($\sigma$)-distribution conditioned to $\{1,2,\ldots\}$, i.e.
\begin{equation}
 \mathbb{P}(K_{r}^{} = n) = \frac{\sigma^n_{}}{n! (\exp (\sigma) - 1)}, \quad n\in \mathbb N.
 \label{bedingtePoissonVtlg}
\end{equation}
We may construct the {\em equilibrium ASG} as in \citet*{Pokalyuk} in  two stages: first take a random path $(K_r)_{-\infty<r<\infty}$, and then fill in the branching and coalescence events, with a random choice of one of the $K_r$ lines at each upward jump, and of one of the $\binom{K_r}{2}$ pairs at each downward jump of $(K_r)$.
Mutation events 
(at rates $\theta \nu_0^{}$ and $\theta \nu_1^{}$)
are superposed on the lines of the (equilibrium) ASG
by Poisson processes with  rates $\theta \nu_0^{}$ and $\theta \nu_1^{}$.
Given the frequency $x$ of the beneficial type at time $0$, one then assigns types to the lines of the ASG in the (forward) time interval $[0,\infty)$ by first drawing the types of the lines at time $0$ independently and identically distributed (i.i.d.) from the weights $(x,1-x)$, and propagates 
the types forward in time, respecting the mutation
events.
In this way, the (backward in time) branching events may now be resolved into the true parent 
and a fictitious parent. 

\begin{figure}[ht]
\psfrag{r}{\small$r$}
\begin{minipage}[t]{0.47\textwidth}
\begin{center}
\includegraphics[width=0.98\textwidth]{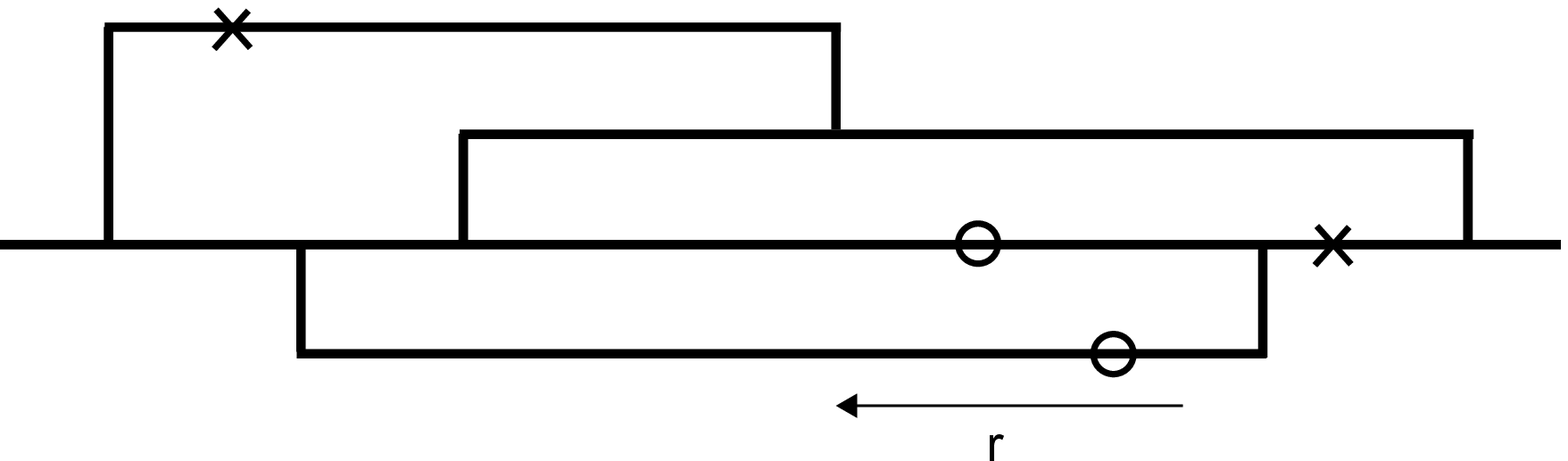}
\end{center}
\end{minipage}
\begin{minipage}[t]{0.47\textwidth}
\begin{center}
\includegraphics[width=0.98\textwidth]{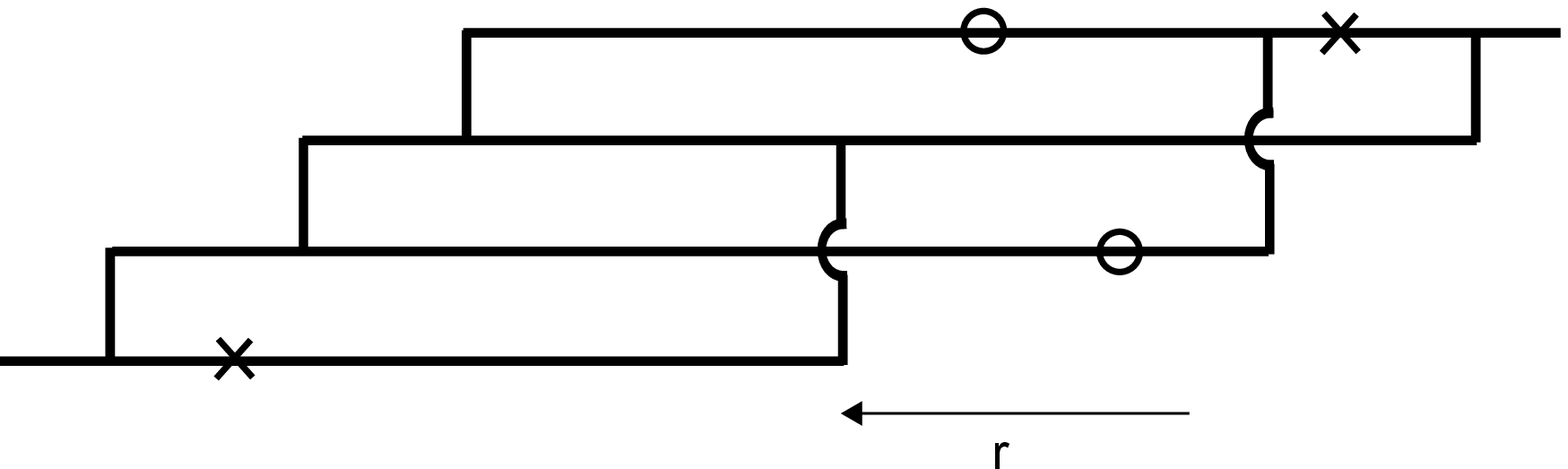}
\end{center}
\end{minipage}
\caption{Different representations of the same ASG realisation with superimposed mutation. All potential ancestors  of the  line next to the top in Fig. \ref{MoranModel} are shown (before resolution into true and fictitious parents).} 
\label{ASGwieNeuhauserKrone}
\end{figure}

Note that 
there are various ways to illustrate the same realisation of the ASG graphically.
See, for instance, Fig.~\ref{ASGwieNeuhauserKrone}, with backward time $r$ running from right to left. The left 
and right panels of Fig.~\ref{ASGwieNeuhauserKrone} represent the same realisation of the ASG, but differ in the 
ordering of the lines.

\subsection{The common ancestor}
\label{sectioncommonancestor}

\begin{figure}[ht]
\begin{center}
\psfrag{t=0}{\Large$t=0$}
\psfrag{t}{\Large$t$}
\psfrag{0}{\Large$0$}
\psfrag{1}{\Large$1$}
\psfrag{s}{\Large$s$}
\psfrag{CA}{\Large CA}
\includegraphics[angle=0, width=0.6\textwidth]{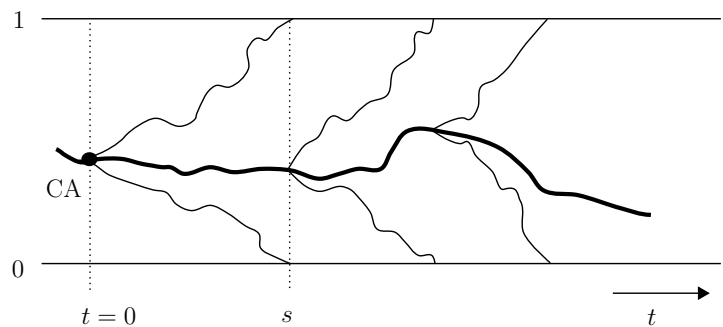}
\caption{The common ancestor at time $t=0$ (CA) is the individual whose progeny 
will eventually fix in the population (at time $s$).}
\label{aline}
\end{center}
\end{figure}

In the population, at any time
$t$, there almost surely exists a unique individual that is, at some time $s > t$,
ancestral to the whole population; cf. Fig.~\ref{aline}. The descendants of this individual become
fixed, and we call it the \textit{common ancestor at time $t$}. The lineage of these
distinguished individuals over time defines the so-called \textit{ancestral}
  (or \textit{immortal}) \textit{line}.

Looking at the population at time $t$, say $t=0$, we are interested in $h(x)$, the probability that the common ancestor is of type $0$, given 
$X_0^{}=x$. Equivalently, one may understand $h(x)$ as the 
probability that the offspring of all type-$0$ individuals 
(regardless of the offspring's types) will ultimately
be ancestral to the entire population,
if $X_0^{}=x$. The probability $h(x)$ does depend on
the type-$0$ frequency $x$ at that time but not on the time itself.
According to previous results by \citet{Fearnhead} and \citet{Taylor}, it reads
\begin{equation}
 h(x) = \sum_{n \geq 0} a_n^{} x (1-x)^n,
 \label{hTaylor}
\end{equation}
where the coefficients $a_n^{}$, $n \geq 0$, are characterised by the recursion
\begin{equation} 
 (n+1+\theta \nu_1^{}) a_{n+1}^{} - (n+1 + \sigma + \theta) a_n^{} + \sigma a_{n-1}^{} =0, \quad  n \geq 1
 \label{recursion}
\end{equation}
under the constraints
\begin{equation}
 a_0 =1, \qquad   \lim_{n\to \infty} \frac{a_{n+1}}{a_n} = 0 .
 \label{initialValues}
\end{equation}
Also, it is shown in \citet{Fearnhead} that \eqref{recursion} and \eqref{initialValues} imply
\begin{equation}\label{tails}
1 = a_0 \ge a_1 \ge \cdots, \qquad \lim_{n\to \infty} a_n = 0.
\end{equation}
These results were reviewed by \citet*{Baake}, and re-obtained with the help of a \emph{descendant process} (forward in time)
by \citet{Kluth}. 

Eq. \eqref{hTaylor} quantifies the bias towards type 0 on the immortal line. The $n=0$ term on the right-hand side of \eqref{hTaylor} is
  $a_0 x=x$, which coincides with the fixation probability in the neutral
  case ($\sigma=0$). Indeed, for $\sigma=0$, we have $a_0=1$, $a_i=0$ for $i \geq 1$ (this is easily seen to satisfy \eqref{hTaylor} and \eqref{recursion}).  For
  $\sigma > 0$, however, all $a_i$ are positive (again by inspection of
\eqref{hTaylor} and \eqref{recursion}), and the terms for $n \geq 1$ in (3)
  quantify the long-term advantage of the favourable type.

In order to get a handle on the representation~\eqref{hTaylor} and the recursion~\eqref{recursion} in terms of the equilibrium ASG,
one observes that the type of the common ancestor at time $t=0$ may be recovered in the following 
way. In the equilibrium ASG marked with the mutation events (as described in Section~\ref{sectioncommonancestor}), assign i.i.d types to the lines at time $0$ and propagate them forward in time, respecting the mutation events. The immortal line is then encoded in the realisation of the marked ASG.

\begin{figure}[ht]
\begin{center}
\psfrag{t}{\Huge$0$}
\psfrag{t1}{\Huge$t_0^{}$}
\psfrag{t+T}{}
\includegraphics[angle=0, width=10cm]{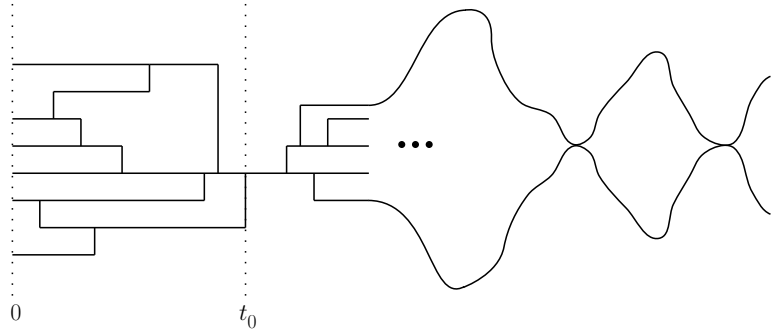}
\caption{A realisation of the equilibrium ASG, with its first bottleneck (after time $0$) at time
 $t_0^{}$.}
\label{ManoBild}
\end{center}
\end{figure}

The event of fixation of the beneficial type is easily described in the case without mutations.
First, recall that, as stated in Section~\ref{sectioncommonancestor}, the number $K_0$ of lines  in the equilibrium ASG at time~$0$ is Poisson distributed with parameter $\sigma$, conditioned to be positive.
Next, observe that, with probability $1$, the equilibrium ASG has bottlenecks, i.e. times at which it consists of a single line.  Let $t_0$ be the smallest among all the non-negative times at which there is a bottleneck (see  Fig.~\ref{ManoBild}). This way,
the unique individual is identified that is the true ancestor of the single individual at 
forward time $t_0^{}$ and, at the same time, of the entire equilibrium ASG at any later time (and ultimately of the entire population). 

As observed by \citet{Mano} and \citet*{Pokalyuk},  type $0$ becomes fixed if and only if the single line at time 
$t_0^{}$ carries type $0$, and this, in turn, happens if and only if at least one ancestral line at time $t=0$ is of type $0$. The latter
probability is $1 - (1-x)^{K_{0}^{}}$, given that the frequency of type-$0$ individuals 
is $x$ at this time. Therefore, with the help of \eqref{bedingtePoissonVtlg}, the fixation probability can be obtained as
\begin{equation}
 h(x)= \mathbb{E} \big (1-(1-x)^{K_{0}^{}} \big )
 = \frac{1}{ \exp (\sigma) - 1} \sum_{n \geq 1} \big (1-(1-x)^n \big ) \frac{\sigma^n}{n!}
 = \frac{1-\exp(- \sigma x)}{1 - \exp(- \sigma)},
 \label{hKimura}
\end{equation}
which coincides with the classical result of \citet{Kimura}. Putting
$\gamma_n^{} := \mathbb{P}(K_{0}^{} > n)$, $n \geq 0$, the left-hand side of 
\eqref{hKimura} may also be expressed as
\begin{equation*}
 h(x) = \sum_{n \geq 1} \big (1-(1-x)^n \big ) \big [ \mathbb{P}(K_{0}^{} \geq n) - \mathbb{P}(K_{0}^{} \geq n+1) \big ]
 = \sum_{n \geq 0} \gamma_n^{} x (1-x)^n ,
\end{equation*}
which is the representation \eqref{hTaylor}. (Indeed, one checks readily that the tail probabilities $\gamma_n$ satisfy the recursion \eqref{recursion} in the case $\theta = 0$.)
The elegance of this approach lies is the fact that one does not need to
know the full representation of the ASG, in particular one does not need to distinguish 
between incoming and continuing branches. As soon as mutations are included, however,
keeping track of the hierarchy of the branches becomes 
a challenge. We thus 
aim at an alternative representation of the ASG that allows 
for an orderly bookkeeping leading to a generalisation of the idea above, and yields a graphical
interpretation of \eqref{hTaylor}-\eqref{initialValues}. This will be achieved in the next three sections.

\section{The ordered ASG}
\label{thronfolgeordnung}
\begin{figure}[ht]
\psfrag{r}{\small$r$}
\begin{center}
\includegraphics[width=0.6\textwidth]{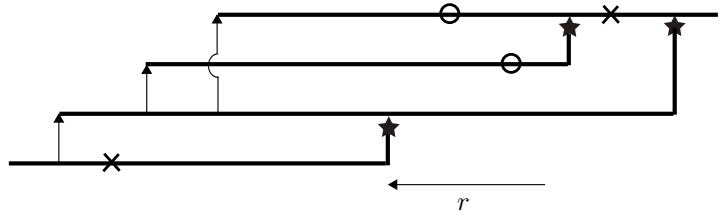}
\end{center}
\caption{The ordered ASG corresponding to Fig.~\ref{ASGwieNeuhauserKrone} or directly deduced from Fig.~\ref{MoranModel}.}
\label{ASGThronfolgeordnung}
\end{figure}

In the previous section, we have reminded ourselves that one may represent the same realisation of one ASG in different ways.
In the following, we propose a construction,
which we call the \textit{ordered ASG}, and which is obtained backwards in time from a given realisation of the ASG as
follows (compare Fig.~\ref{ASGThronfolgeordnung}).
\begin{itemize}
	\item \textit{Coalescence:} Each coalescence event is represented by a (neutral) arrow pointing from the lower participating line to the upper one. The (single) parental line continues back in time from the lower branch.
	\item \textit{Branching:} A selective arrow with star-shaped head is pointed towards the splitting line at a branching event. 
	The incoming branch is always placed directly beneath the continuing branch at the tail of the arrow;
	in particular, there are no lines between incoming and continuing branch at the time
	of the branching event.
	\item \textit{Mutation:} Mutations,  symbolised here by circles and crosses, occur along the lines as in the original ASG. 
\end{itemize}
The ordered ASG corresponding to both representations in Fig.~\ref{ASGwieNeuhauserKrone}
is shown in Fig.~\ref{ASGThronfolgeordnung}.

\section{The lookdown ASG}
\label{pfeileundsternchenASG}
To each point in the ordered ASG, let us introduce two coordinates: its  \textit{time} and its \textit{level}, the latter being an element of $\{1,2,\ldots\}$ which coincides with the number of lines in the ASG. 
Since this is in close analogy to ideas known from the lookdown processes
by \citet*{DonnellyKurtz99,DonnellyKurtz99b}, 
we  call this construction the \textit{lookdown ASG (LD-ASG)}.
It can be obtained backwards in time from a given realisation of the ordered ASG, or it may as well be constructed in distribution via Poissonian elements representing coalescence, branching, and mutation. The two possibilities are described in Sections~\ref{sec:fromreal} and \ref{sec4_2}, respectively. 

\subsection{Construction from a given realisation of the ordered ASG}
\label{sec:fromreal}

\begin{figure}[ht]
\begin{minipage}[t]{0.45\textwidth}
\begin{center}
\includegraphics[width=0.4\textwidth]{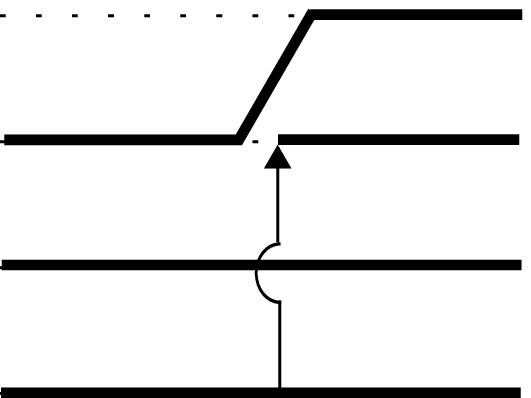}
\end{center}
\end{minipage}
\begin{minipage}[t]{0.45\textwidth}
\begin{center}
\includegraphics[width=0.4\textwidth]{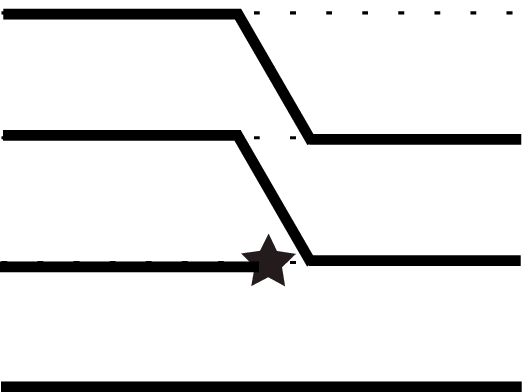}
\end{center}
\end{minipage}
\caption{Coalescence (left) and branching event (right) in the LD-ASG.}
\label{vierelementebilder}
\end{figure}

Backwards in time, the realisation of the LD-ASG corresponding to a given realisation of the ordered ASG is obtained in the following way. Start with all $n$ individuals (respectively lines) that are present  in the (ordered) ASG and place
	them at levels $1$ to $n$ by adopting their vertical order from the ordered ASG. Then let the following events happen (backward in time):
\begin{itemize}
	\item \textit{Coalescence:}
	Coalescence events between levels $i$ and $j>i$ are treated the same way as in the ordered ASG: The remaining branch continues at level $i$. In addition, all lines at levels $k>j$ are shifted one level downwards to $k-1$ (cf. Fig.~\ref{vierelementebilder}, left).
	\item  \textit{Branching:}
	A selective arrow with star-shaped head in the ordered ASG is translated into a star at the level $i$ of the branching line. The incoming branch emanates out of the star at the same level and all lines at levels $k\geq i$ are pushed one level upwards to $k+1$. In particular, the continuing branch is shifted	to level $i+1$ (cf. Fig.~\ref{vierelementebilder}, right).
		\item \textit{Mutation:}  Mutations (symbolised again as circles and crosses) are taken from the ordered ASG.
\end{itemize}

\begin{figure}[ht]
\psfrag{r}{\small$r$}
\psfrag{0}{\small$0$}
\psfrag{1}{\small$1$}
\psfrag{2}{\small$2$}
\psfrag{3}{\small$3$}
\psfrag{4}{\small$4$}
\psfrag{5}{\small$5$}
\begin{center}
\includegraphics[width=0.6\textwidth]{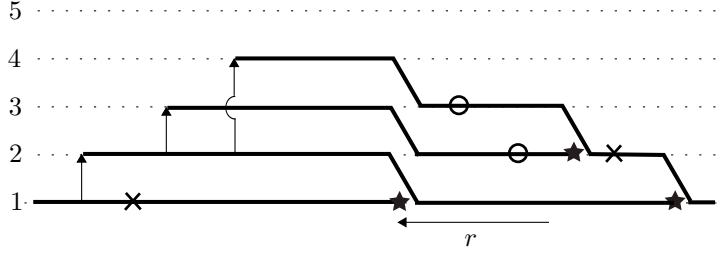}
\end{center}
\caption{LD-ASG corresponding to Fig.~\ref{ASGThronfolgeordnung}. Levels are numbered from bottom to top. }
\label{ASGLookdown}
\end{figure}

Fig.~\ref{ASGLookdown} gives a realisation that corresponds to the realisation of the ordered ASG in Fig.~\ref{ASGThronfolgeordnung}.
Note that we obviously have a bijection between realisations of the ordered ASG and the LD-ASG and that 
the LD-ASG is just a neat arrangement of the ordered ASG. 

\subsection{Construction from elements of Poisson point processes}\label{sec4_2}

The LD-ASG may, in distribution, as well be constructed backwards in time via  the elements `arrows', `stars', `circles', and `crosses' arising as representations of independent Poisson point processes:

\begin{itemize}
	\item \textit{Coalescence:}
	 For each ordered pair of levels $(i,j)$, where $ i<j$ and level $j$
	is occupied by a line, \textit{arrows}
	from level $i$ to $j$ emerge independently according to a Poisson point process $\Gamma^{\uparrow}_{ij}$ at rate $2$.
	An arrow from $i$ to $j$ is understood as a coalescence of the lines at levels $i$ and $j$ to a single line on level $i$.
	In addition, all lines at levels $k > j$ are shifted one level downwards to $k-1$ (cf. Fig.~\ref{vierelementebilder}, left).  
	\item  \textit{Branching:}
	On each occupied level $i$ \textit{stars} appear according to independent Poisson point processes $\Gamma^{\ast}_i$ at rate $\sigma$.
	A star at level $i$ indicates a branching event, where a new line, namely the incoming branch, is inserted at level $i$ and all lines at levels $k\geq i$ 
	are pushed one level upwards to $k+1$. In particular, the continuing branch is shifted
	to level $i+1$ (cf. Fig.~\ref{vierelementebilder}, right).
	\item \textit{Mutation:}  Mutations to type $0$ and type $1$, i.e. circles and crosses, occur via independent Poisson point processes $\Gamma^{\circ}_i$ at rate $\theta\nu_0^{}$ and $\Gamma^{\times}_i$ at rate $\theta\nu_1$, respectively, on each occupied level $i$.
\end{itemize}

The independent superposition of these Poisson point processes and their effects on the lines characterises the LD-ASG.
Recall that $(K_{r}^{})_{r \in \mathbb{R}}^{}$ is the line counting process of the (ordered) ASG and thus $K_r$ is also the highest occupied level of the LD-ASG at time $r$. It evolves backwards in time with transition rates given by \eqref{ratesASGNKbackwards}.


Note  that, although we will
ultimately rely on the ASG in equilibrium only, neither the  ordering of the
ASG nor the LD-ASG construction are restricted to the equilibrium situation.
The equilibrium comes back in when we search for the immortal line, which
will be done next.

\subsection{The immortal line in the LD-ASG in the case without mutations}
\label{Thronfolgeordnung}

\begin{figure}[ht]
\psfrag{r}{\small$r$}
\psfrag{t0}{\small$t_0$}
\psfrag{0}{\small$0$}
\psfrag{1}{\small$1$}
\psfrag{2}{\small$2$}
\psfrag{3}{\small$3$}
\psfrag{4}{\small$4$}
\psfrag{5}{\small$5$}
\begin{center}
\includegraphics[width=0.6\textwidth]{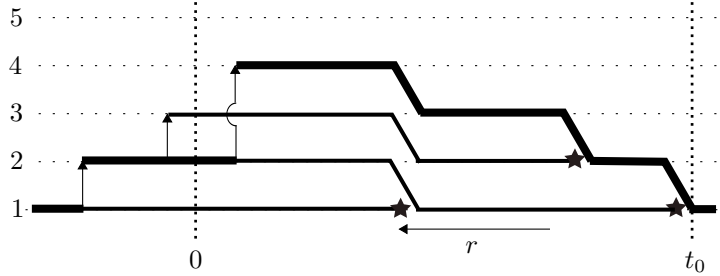}
\end{center}
\caption{LD-ASG (without mutations) corresponding to Fig. \ref{ASGThronfolgeordnung}. The immune line is marked bold.}
\label{ASGLookdown2}
\end{figure}

We consider a realisation $\mathcal G$ of the equilibrium LD-ASG, write $K_r$ for its highest occupied level at (backward) time $r$, and again write
 \begin{equation}\label{t0}
 t_0 = t_0(\mathcal G) := -\sup\{r\le 0: K_r = 1\}
 \end{equation}
 for the smallest (forward) time at which $\mathcal G$ has a `bottleneck', see Fig.~\ref{ASGLookdown2}.

\begin{figure}[ht]
\begin{center}
\psfrag{0}{\Large$0$}
\psfrag{1}{\Large$1$}
\psfrag{r}{\LARGE$r$}
\psfrag{0/1}{\Large$0/1$}
\includegraphics[angle=0, width=0.45\textwidth]{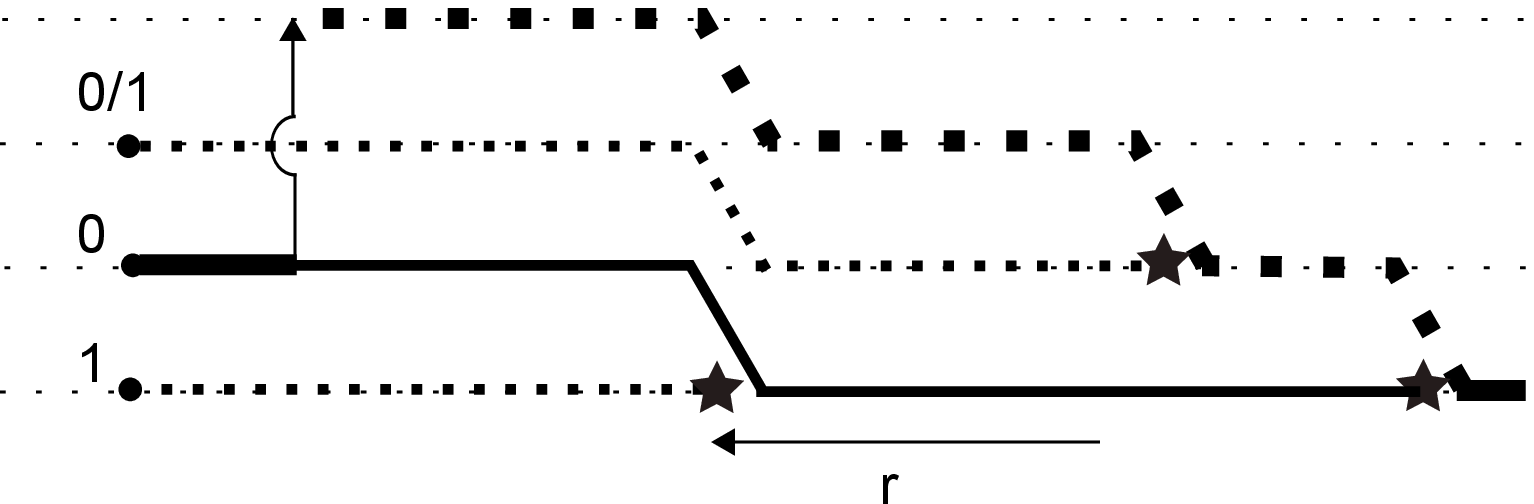}
\hspace{3ex}
\includegraphics[angle=0, width=0.45\textwidth]{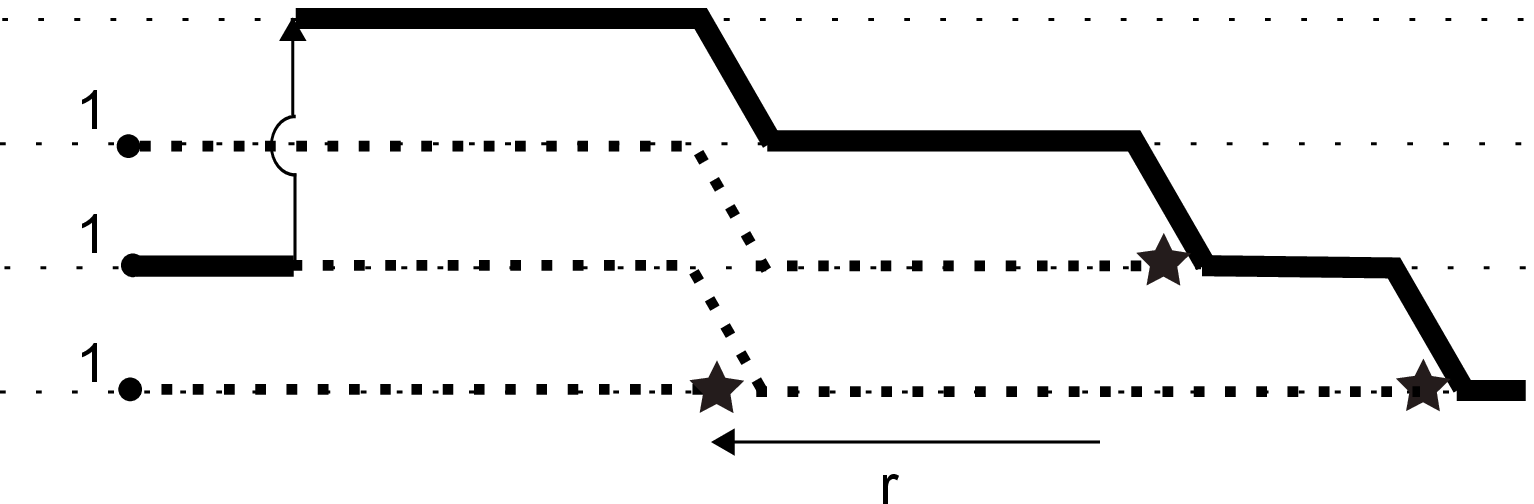}
\caption{LD-ASG with types. The level of the immortal line (solid) starting out from time $0$ depends on the type assignment at time $0$. The immune line is marked bold (and is the immortal line in the right picture).}
\label{lookdownimmortal}
\end{center}
\end{figure}

The level of the  immortal line at time $0$ does not only depend  on $\mathcal G$, but also on the types $I^1,\ldots, I^{K_0} \in \{0,1\}$  that are assigned to the levels $1, 2, \ldots, K_0$ at time $0$. We now define a distinguished line which we call the {\em immune line}. The reason for this naming will become clear in the next section: the immune line will be  exempt from the pruning.
\begin{defi} \label{def1} At any given time, the {\em immune line} is the line that will be immortal if all lines at that time are of type 1.
\end{defi}
The following is immediate from the construction of the LD-ASG: back from each bottleneck of $\mathcal G$, the immune line goes up one level at each branching event that happens at a level smaller or equal to its current level, and follows the coalescence events in a lookdown manner, see the bold line in the right panel of Fig.~\ref{lookdownimmortal}. In particular, the immune line follows the continuing branch whenever it is hit by a branching event at its current level.

The next proposition is illustrated by Fig.~\ref{lookdownimmortal}.
\begin{proposition}\label{remarknomut}
 In the absence of mutations, for almost every realisation $\mathcal G$ of  the equilibrium LD-ASG with types assigned at time $0$,  the level of the immortal line  at time $0$ is either the
lowest type-$0$ level at time $0$  or, if all lines at time $0$ are of type $1$, it is the level of the immune line at time $0$.
\label{Thronfolgelemma}
\end{proposition}

\emph{Proof.}
 We proceed by induction along the Poissonian elements ``branching'' and ``coalescence'' described in Sec. \ref{sec4_2}, backwards from $t_0$, the first time after time $0$ at which the number of lines in $\mathcal G$ is one. Let $t_k < 0 < t_{k-1} < \cdots < t_0$ be the times at which these elements occur (note that for almost every realisation $\mathcal G$ the number $k$ is finite, and all the $t_j$ are distinct) and choose times $0=:s_k < \cdots < s_0$ with $t_j<s_j <t_{j-1}$, $1\le j\le k$. We claim that for all $j=1,\ldots, k$, when assigning types at time $s_j$, the level of the immortal line  at time $s_j$ is either the
lowest type-$0$ level at time $s_j$  or, if all lines at time $s_j$ are of type $1$, it is the level of the immune line at time $s_j$. 

\begin{figure}[ht]
\begin{center}
\psfrag{0}{\Large$0$}
\psfrag{1}{\Large$1$}
\includegraphics[angle=0, width=0.6\textwidth]{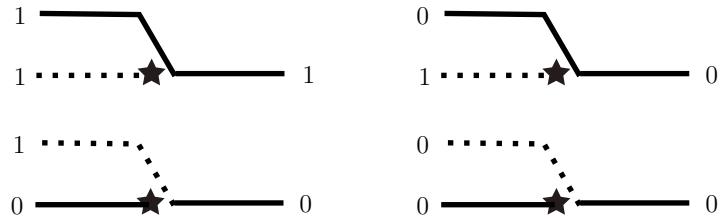}
\caption{Branching event in the LD-ASG. The four possible combinations of types are shown (in analogy with Fig.~\ref{IC}). The parental branch (bold line) is the incoming one (upper two diagrams) if it is of type $0$, and the continuing one (lower two diagrams) if the incoming branch is of type $1$.}
\label{lookdownancestor}
\end{center}
\end{figure}

The assertion is obvious for $j=1$, since by assumption no event has happened between $s_1$ and $t_0$. Now consider the induction step from $j$ to $j+1$. If all lines at time $s_{j+1}$ are assigned type 1, then the level of the immortal line at time $s_{j+1}$ is by definition that of the immune line at  time $s_{j+1}$. Now assume that at least one line at time $s_{j+1}$ is assigned type~0. If the event at time $t_{j}$ was a coalescence (such as the leftmost event in Fig.~\ref{lookdownimmortal}), then by the induction assumption the level of the immortal line  at time $s_{j+1}$ is  the
lowest type-$0$ level at time $s_{j+1}$ (this is because the types are
propagated along the lines, in particular along the line at the tail
of the arrow). 
%
%
%
If, on the other hand, the event at time $t_{j}$ was a branching, then again by the induction assumption and by the ``pecking order''  illustrated in  Fig.~\ref{lookdownancestor}, the level of the immortal line  at time $s_{j+1}$ is  the
lowest type-$0$ level at time $s_{j+1}$. This completes the proof of the proposition. \qed

\medskip
Proposition~\ref{Thronfolgelemma} specifies the immortal line in the case of selection only.
The aim in the next section is to establish an analogous statement in the case of selection
\textit{and} mutation.

\section{The pruned equilibrium LD-ASG and the CAT distribution}
\label{mutations}

We now consider the equilibrium LD-ASG marked with the mutation events. Working backwards from the bottleneck time $t_0$ (see Eq.~\eqref{t0} and Figs.~\ref{ASGLookdown} and \ref{ASGLookdown2}), we see that the mutation events that occur along the lines may eliminate some of them as candidates for being the immortal line. Cutting away certain branches that carry no information has been used, explicitly or implicitly, in various investigations of the ASG (e.g. by \citet{Slade}, \citet{Fearnhead}, \citet*{AthreyaSwart}, and \citet*{EtheridgeGriffiths}), but
our construction requires a specific {\em pruning procedure} which we now describe.

\subsection{Pruning the LD-ASG}\label{PrASG}
\begin{figure}[ht]
\begin{minipage}[t]{0.45\textwidth}
\begin{center}
\includegraphics[width=0.4\textwidth]{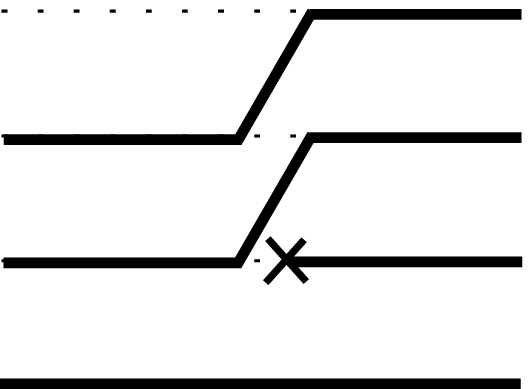}
\end{center}
\end{minipage}
\begin{minipage}[t]{0.45\textwidth}
\begin{center}
\includegraphics[width=0.4\textwidth]{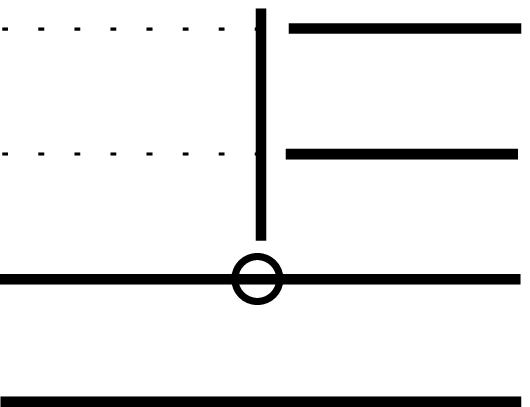}
\end{center}
\end{minipage}
\caption{Pruning procedure in the LD-ASG due to deleterious  (left) and beneficial (right) mutations that appear on lines that are not immune.}
\label{KreuzchenKringelLookdownASG}
\end{figure}

\begin{figure}[ht]
\begin{center}
\psfrag{r}{\small$r$}
\psfrag{0}{\small$0$}
\psfrag{1}{\small$1$}
\psfrag{2}{\small$2$}
\psfrag{3}{\small$3$}
\psfrag{4}{\small$4$}
\psfrag{5}{\small$5$}
\includegraphics[width=0.8\textwidth]{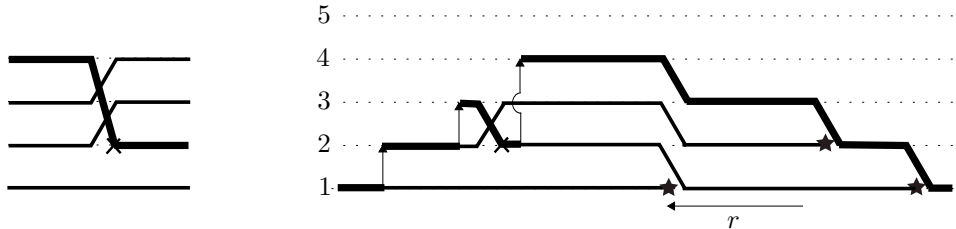}
\end{center}
\caption{Relocation procedure in the LD-ASG due to deleterious mutations on the immune line (bold).}
\label{KreuzchenimmuneLookdownASG}
\end{figure}

In addition to the events {\em branching} and {\em coalescence} we now have the {\em deleterious mutations} and the {\em beneficial mutations}. As in the proof of Proposition \ref{remarknomut}, let $t_k < 0 < t_{k-1} < \cdots < t_0$ be the times at which all these events occur, and choose times $0=:s_k < \cdots < s_0$ with $t_j<s_j <t_{j-1}$, $1\le j\le k$. Assume  branching and coalescence events (but no mutation events) happen at the times $t_1, \ldots, t_{i-1}$, and a mutation event happens at time $t_i$. Recall from Definition 1 that at any given time the immune line is the line that will be immortal if all lines at that time are of type 1; but now the rule for the immune line must be adapted due to the
impact of mutations.
First consider the case in which our first mutation is deleterious (symbolised by a cross). Since there is no mutation between times $t_i$ and $t_0$, Proposition \ref{remarknomut} applies (with time $0$ replaced by time $s_{i+1}$), showing that the line that is hit by the deleterious mutation at time $t_i$ cannot be the immortal one unless it is the immune line.  In our search  for the true ancestor of the line that goes back from time $t_0$ we can therefore {\em erase} the line segment to the left of $t_i$, unless the line in question is the immune one; all lines above the one that is erased slide down one level to fill the space, see Fig.~\ref{KreuzchenKringelLookdownASG} (left). If the immune line is hit by a deleterious mutation at $t_i$, it is the immortal line at time $s_{i+1}$ if and only if all the other lines at time $s_{i+1}$ are of type $1$. In order to tie in with our picture that the level of the immortal line at any time is the lowest level that is assigned type $0$ (given there is at least one lineage at this time that is assigned type $0$), we {\em relocate} our mutated immune line  to the currently highest level of the LD-ASG at time $t_i$, whereby the levels of all the other lines that were above the immune line at time $s_i$ are shifted down by 1 (compare Fig. \ref{KreuzchenimmuneLookdownASG}).

Next consider the case in which the mutation occurring at time $t_i$ is beneficial (symbolised by a circle)  and happens at level $\ell$, say.  Then, again appealing to Proposition~\ref{remarknomut}, we see that none of the lines that occupy  levels $>\ell$ at time $t_i$ can be parental to the single line that exists at time $t_0$.  We can therefore erase all these lines from the list of candidates for the ancestors. We indicate this by inserting a barrier of infinite height  above the circle, see Fig.~\ref{KreuzchenKringelLookdownASG} (right).
If all lines at time $s_{i+1}$ are assigned type 1, then the line on level $\ell$  becomes the only one that carries type 0 at time $s_i$ and therefore 
is  immortal. Thus, the immune line is relocated  to level $\ell$ at time $t_i$.

\medskip
Proceeding to the next mutation event on the remaining lines back from time $t_i$ (which happens at time $t_m$ at level $\ell'$, say), we can iterate this procedure: if the mutation is deleterious, the line is killed unless it is the immune one. If the immune line is hit by a deleterious mutation, then it is relocated to the currently highest level of the LD-ASG. If the mutation is beneficial, all the lines at higher levels are killed, with the line starting back from $t_m$ at level $\ell'$ being declared the new immune line. 


Having worked back to $t=0$,  we arrive at  the \textit{pruned LD-ASG} between times $0$ and $t_0$. 
Note that except for the immune line, due to the pruning procedure there are no mutations on any line of  the \textit{pruned LD-ASG}.
\begin{figure}[ht]
\psfrag{r}{\small$r$}
\psfrag{0}{\small$0$}
\psfrag{1}{\small$1$}
\psfrag{2}{\small$2$}
\psfrag{3}{\small$3$}
\psfrag{4}{\small$4$}
\psfrag{5}{\small$5$}
\begin{center}
\includegraphics[width=0.6\textwidth]{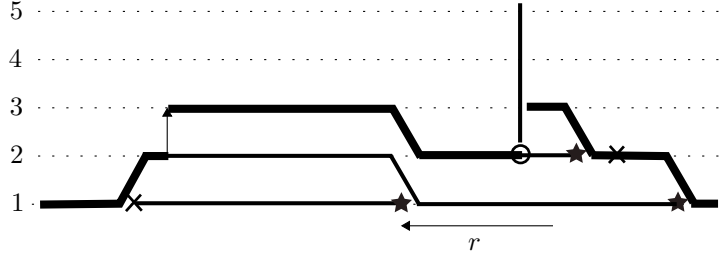}
\end{center}
\caption{Pruned LD-ASG derived from Fig.~\ref{ASGLookdown}. The immune line is marked bold.}
\label{ASGBeispielpruned}
\end{figure}	
In other words, each line present at time $0$ is either the immune line, or has no mutations on it between times $0$ and $\zeta$, where $\zeta$ is the time when that line was incoming to a branching event with the immune line. Note also that beneficial mutations can only be present on the current top level of the pruned LD-ASG. \\

As in Section~\ref{sec4_2} we can construct the  pruned LD-ASG (together with the level of the immune line) backwards in time in a Markovian way, using  the Poisson processes $\Gamma^{\uparrow}_{ij}$
and $\Gamma^{\ast}_i$ (for all occupied levels $i$ and $j$, cf. Fig.~\ref{vierelementebilder}), and  $\Gamma_i^{\times}$ and
$\Gamma_i^{\circ}$, where the pruning procedure is applied as described above (cf. Fig.~\ref{KreuzchenKringelLookdownASG}).
Fig.~\ref{ASGBeispielpruned} gives a realisation that corresponds to the realisation
of the LD-ASG in Fig.~\ref{ASGLookdown}.

\subsection{The line-counting process of the pruned LD-ASG}
The construction of the previous subsection shows that the process $(L_{r}^{})_{r \in \mathbb{R}}^{}$,
where $L_{r}^{}$ is the \textit{level of the top line} (which coincides with the number of lines) at the backward time $r=-t$, evolves with 
transition rates
\begin{equation}
\begin{split}
&{q}_{L}^{}(n,n-1) = n(n-1)  + (n-1)\theta\nu_1 + \theta\nu_0 ,\\
&{q}_{L}^{}(n,n+1) = n\sigma, \\
&{q}_{L}^{}(n,n-\ell) = \theta\nu_0 ,  \hspace{25ex} 2 \leq \ell < n, \quad n \in \mathbb{N} .
\end{split}
\label{ratesprunedASGbackwards}
\end{equation}
In words, when the top level is currently $n$, it decreases by one 
when either a coalescence event
happens between any pair of lines (rate $n (n-1)$), 
or one of the $n-1$ lines that are not immune experiences a deleterious mutation
(rate  $(n-1)\theta\nu_1$), or line $n-1$ experiences a beneficial
mutation (rate $\theta\nu_0$). The top level increases by one when one
of the lines branches (rate $n \sigma$). It decreases by $\ell$,
$2 \leq \ell < n$, when level $n-\ell$ experiences a beneficial
mutation (rate $\theta\nu_0$) .

\begin{remark}	\label{dom}
The process $L$ is stochastically dominated by the  process $K$ (the highest level of the unpruned LD-ASG). In fact, using the above-described pruning procedure in a time-stationary picture  between all the bottlenecks of the equilibrium ASG line counting process $K= (K_{r}^{})_{r \in \mathbb{R}}^{}$, we obtain that $L_r \leq K_r$ for all $r \in \mathbb R$.\end{remark}
The stochastic dynamics induced by \eqref{ratesprunedASGbackwards} thus has a unique equilibrium distribution, which we denote by $\rho$. In the following, let $L= (L_r)_{r \in \mathbb{R}}^{}$ be the time-stationary process with jump rates~\eqref{ratesprunedASGbackwards}. We then have
\begin{equation}\label{L0}
\rho_n = \mathbb P(L_0 =n), \quad n\in \mathbb N,
\end{equation}
and $\rho = (\rho_n)$ is the probability vector obeying
\begin{equation}\label{equrho}
\rho Q = 0,
\end{equation}
with $Q$ being the generator matrix determined by the jump rates~\eqref{ratesprunedASGbackwards}.

\subsection{The type of the immortal line in the pruned LD-ASG}
We will now show that the type of the immortal line at time $0$ is determined by the type configuration assigned at time $0$  in a way quite similar to  the case without mutations. 
\begin{theorem} \label{typeimm}
For almost every realisation $\mathcal G$ of  the pruned equilibrium LD-ASG with types assigned at time $0$,  the level of the immortal line  at time $0$ is either the
lowest type-$0$ level at time $0$  or, if all lines at time $0$ are of type $1$, it is the level of the immune line at time $0$.
In particular, the immortal line is of type $1$ at time  $0$ if and only if all lines in $\mathcal G$ at time $0$ are assigned the type $1$.
\label{immortallineinlookdownASG}
\end{theorem}
\emph{Proof.}
 We proceed by induction along the Poissonian elements described in Sec.~\ref{PrASG} backwards from $t_0$, the first time after time $0$ at which the number of lines in $\mathcal G$ is one. As in Sec.~\ref{PrASG}, let $t_k < 0 < t_{k-1} < \cdots < t_0$ be the times at which these elements occur, and choose times $0=:s_k < \cdots < s_0$ with $t_j<s_j <t_{j-1}$, $1\le j\le k$. We will prove that for all $j=1,\ldots, k$  the level of the immortal line  at time $s_j$ is either the
lowest type-$0$ level at time $0$  or, if all lines at time $s_j$ are of type $1$, it is the level of the immune line at time $s_j$. 

If the event that occurs at $t_j$ is a branching or a coalescence, then the induction step from $j$ to $j+1$ is precisely as in the proof of Proposition \ref{remarknomut}.
If the event at time $t_j$ is a deleterious mutation, then we distinguish two cases. In the first case, assume that this deleterious mutation happens on the immune line. Then, according to the rule prescribed in Sec.~\ref{PrASG}, the immune line is relocated to the top level of $\mathcal G$ at time $t_j$ (compare Fig. \ref{KreuzchenimmuneLookdownASG}). If there is a line that is assigned type $0$ at time $s_{j+1}$, then the immortal line at time $s_{j+1}$  is found at the lowest type-$0$ level.
(In particular, if the uppermost line at time $s_{j+1}$ is assigned type $0$, then it is the immortal line  if and only if all other  lines at time $s_{j+1}$ are assigned type $1$. This is due to   the deleterious mutation at time $t_j$ and the relocation to the top level.) In the second case assume that the deleterious mutation happens on  a line different from the immune one. Then the number of lines  at time $s_{j+1}$ is one less than the number of lines at time $s_j$; more specifically, all the lines from time $s_{j+1}$ can be found in the same order also at time $s_j$, and in addition at time $s_j$ there is one line carrying type $1$ which cannot be ancestral since it is not the immune line. Thus the induction assumption from time $s_j$ carries over to give the required assertion for time $s_{j+1}$.

Finally, assume that the event at time $t_j$ is a beneficial mutation. In this case, if at time $s_{j+1}$ type 0 is assigned to a level $\ell$ that is occupied by one of the  lines that remain after the pruning at time $t_j$, and if all the levels below $\ell$ are assigned type $1$, then we can infer from the induction assumption that the line at level $\ell$ at time $s_{j+1}$ is the immortal one. On the other hand, if  all the lines that remain at time $s_{j+1}$ are assigned type $1$, then, because of the beneficial mutation at time $t_j$, the top line at time $s_{j+1}$ is the immortal one, and due to our relocation rule this is also the immune line at time $s_{j+1}$. Thus, the induction step is completed, and the theorem is proved.   \qed

\subsection{The CAT distribution via the pruned equilibrium LD-ASG} 
With the help of Theorem~\ref{immortallineinlookdownASG}, it is now possible to provide an interpretation of the probability $h(x)$ that the common ancestor is of type $0$, given that the frequency of the beneficial type at time $0$ is $x$.
\begin{theorem}\label{theorem_h}
Given the frequency of the beneficial type at time $0$ is $x$, the probability that the common ancestor at time $0$ is of beneficial type is
\begin{equation}\label{hLBKW}
h(x)= \sum_{n\geq 1} x(1-x)^{n-1}\mathbb P(L_0 \ge n),
\end{equation}
where $L_0$ is the number of lines at time $0$  in the time-stationary pruned LD-ASG, see formula~\eqref{L0}. \end{theorem}
{\em Proof.}
Let  $I^k \in \left\{0,1\right\}$ be the type that is assigned to the individual at level $k \in \left\{1, \ldots, L_{0}^{}\right\}$ in the pruned equilibrium LD-ASG  at time $0$. According to Theorem \ref{typeimm}, the event that the common ancestor at time $0$ is of type $0$ equals the event that at least one of the $I^k$,  $k \in \left\{1, \ldots, L_{0}^{}\right\}$, is~$0$.
Conditional on the initial frequency of the beneficial type being $x$, these types are assigned in an i.i.d. manner  with $\mathbb{P}(I^k=0)=x$. 
The quantity $h(x)$ thus is the probability that at least one of a random number of i.i.d. trials is a success, where the success probability is $x$ in a random number of trials $L_0$ (which is independent of the Bernoulli sequence with parameter $x$). A decomposition of $h(x)$ according to the first level which is occupied by type~$0$ yields 

\begin{equation}\label{decomp}
\begin{split}
h(x) &= \sum_{n\geq 1} \mathbb{P} (I^{n} = 0,\ I^{k}=1\ \forall k<n, \ L_{0}^{}\ge n) \\
&=  \sum_{n\geq 1} \mathbb{P} (I^{n} = 0,\ I^{k}=1\ \forall k< n) \ \mathbb{P}(L_{0}^{}\ge n).
\end{split}
\end{equation} 
The right hand side of \eqref{decomp} equals that of \eqref{hLBKW}, which completes the proof of the theorem. \qed
\\\\
To compare  \eqref{hLBKW} with \eqref{hTaylor}, we rewrite its  right-hand side  as $\sum_{n\geq 0} x(1-x)^{n}\mathbb P(L_0 \ge n+1)$.  It is then clear from the comparison with \eqref{hTaylor} that the tail probabilities $\alpha_{n}:=  \mathbb P(L_0 > n), n \ge 0$, agree with Fearnhead's coefficients $a_n$. They must therefore obey the recursion \eqref{recursion}. The proof of the following proposition gives a direct argument for this.
\begin{proposition}\label{proprec}
The tail probabilities $\alpha_{n} = \mathbb P(L_0 > n), \, n\ge 0$,  obey the recursion \eqref{recursion}.
\end{proposition}
{\em Proof:}
Let  $\rho = (\rho_n)$ be  the probability vector determined by \eqref{equrho}. We then have
\begin{equation}\label{tailprob}
\alpha_{n} = \sum_{i > n} \rho_{i},  \qquad  n \in \mathbb{N}_0\ .
\end{equation}
 For $n \geq 2$, the $n^{th}$ entry of the vector $\rho Q$ is
\begin{align*}
(\rho Q)_n  & = \rho_{n-1} {q}_L(n-1,n) + \rho_{n+1} {q}_L(n+1,n) + \sum_{j \geq n+2} \rho_{j} {q}_L(j,n)\\
& \qquad - \rho_{n} \left[{q}_L(n,n-1) + {q}_L(n,n+1) +\sum_{\ell =0}^{n-2}{q}_L(n,\ell) \right].
\end{align*} 
Thus, plugging in the jump rates~\eqref{ratesprunedASGbackwards}, Eq. \eqref{equrho} is equivalent to 
\vspace{0.7cm}
\begin{align*}
0 & = \rho_{n-1} (n-1) \sigma + \rho_{n+1} \left[n(n+1) + n\theta\nu_1 \right] + \theta\nu_0 \sum_{j \geq n+1} \rho_{j}\\
& \qquad - \rho_{n} \left[n(n-1)+(n-1)\theta+n\sigma \right] ,\quad  n \geq 2 .
\end{align*} 
Writing this in terms of the tail probabilities~\eqref{tailprob}, rearranging terms, and shifting the index, we obtain
\begin{align*}
0 & = n\left\{-\alpha_n\left[n+1+\theta + \sigma \right] + \alpha_{n+1}\left[n+1+\theta\nu_1\right]+\alpha_{n-1}\sigma \right\}\\
& \qquad +\left(n+1\right)\left\{\alpha_{n+1}\left[n+2+\theta+\sigma\right] - \alpha_{n+2}\left[n+2+\theta\nu_1\right] - \alpha_n \sigma \right\}, \quad n \geq 1,
\end{align*} 
which we abbreviate by
$$n(\alpha F)_n = (n+1)(\alpha F)_{n+1}$$
with the (tridiagonal) matrix $F$ that appears in the recursion \eqref{recursion}. In view of these equalities, the proposition is proved if we can show that
$\lim_{n\to \infty} n(\alpha F)_n = 0$, or, even better, that
\begin{equation}\label{convinf}
\lim_{n\to \infty} n^2\alpha_n = 0.
\end{equation}
To see \eqref{convinf}, recall that as stated in Remark~\ref{dom},  $L_0$ is stochastically dominated by the number $K_0$ of lines in the equilibrium ASG, which has distribution \eqref{bedingtePoissonVtlg}. In particular, $L_0$ has a finite third moment. From this, \eqref{convinf} is immediate since for any non-negative integer-valued random variable $\xi$ one has $\mathbb E[\xi(\xi-1)(\xi-2)] = 3 \sum_{n=0}^\infty n(n-1) \mathbb P(\xi > n)$.  Thus, Proposition~\ref{proprec} is proved. \qed
\\\\
The proof of Proposition  \ref{proprec} allows us to conclude the following
\begin{cor}\label{corFearn}
The solution $(a_n)$ of \eqref{recursion} and \eqref{initialValues} is also characterised by \eqref{recursion} together with the constraint \eqref{tails}.
\end{cor}
{\em Proof:} The constraint \eqref{tails} ensures that $r_n := a_{n-1}-a_n, \, n \ge 1$, $r_n$ are probability weights on $\mathbb N$. Starting from \eqref{recursion} and  working  back in the proof of Proposition \ref{proprec} we arrive at Eq. \eqref{equrho}, which (by irreducibility and recurrence of $Q$) has a unique solution among all the probability vectors on $\mathbb N$. This shows that the solution of the recursion \eqref{recursion} is unique also under the constraint~\eqref{tails}. \qed
\\\\
It is worth noting that property \eqref{initialValues} must hold for the tail probabilities  \linebreak $\alpha_{n} = \mathbb P(L_0 > n),$ \ $n\ge 0$, as well since they agree with the $a_n$. This translates into an assertion on the asymptotics of the hazard function of $L_0$:
$$\lim_{n\to \infty} \mathbb P(L_0=n+1| L_0>n) = 1- \lim_{n\to \infty} \mathbb P(L_0>n+1| L_0>n) =1- \lim_{n\to \infty} \frac{\alpha_{n+1}}{\alpha_n} = 1.$$

\section{Monotonicities in the model parameters}
\label{monotonicities}

\begin{figure}[htdp] 
\begin{center}
\begin{minipage}[t]{0.8\textwidth}
\includegraphics[width=\textwidth]{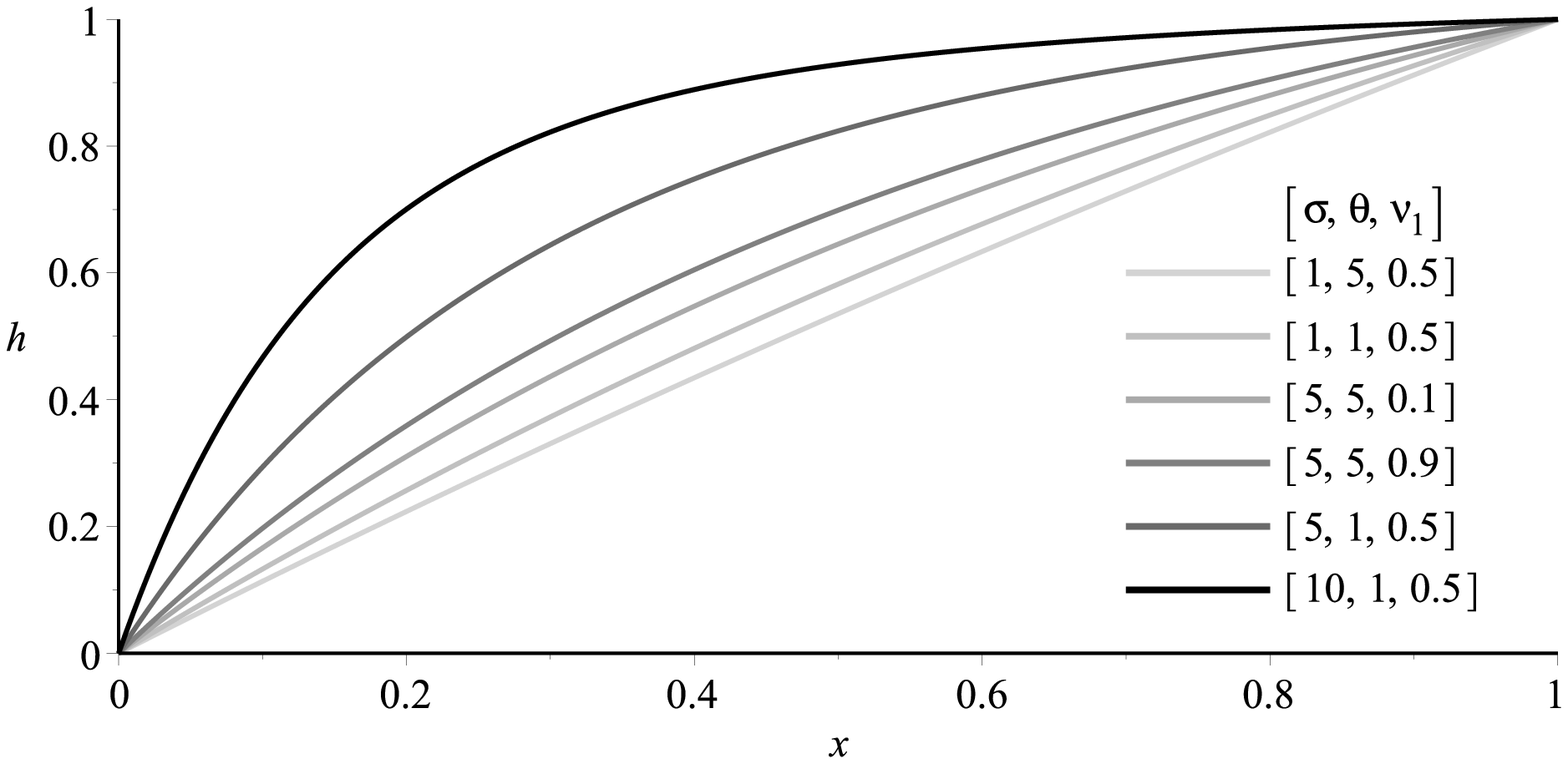}
\end{minipage}
\begin{minipage}[t]{0.8\textwidth}
\includegraphics[width=\textwidth]{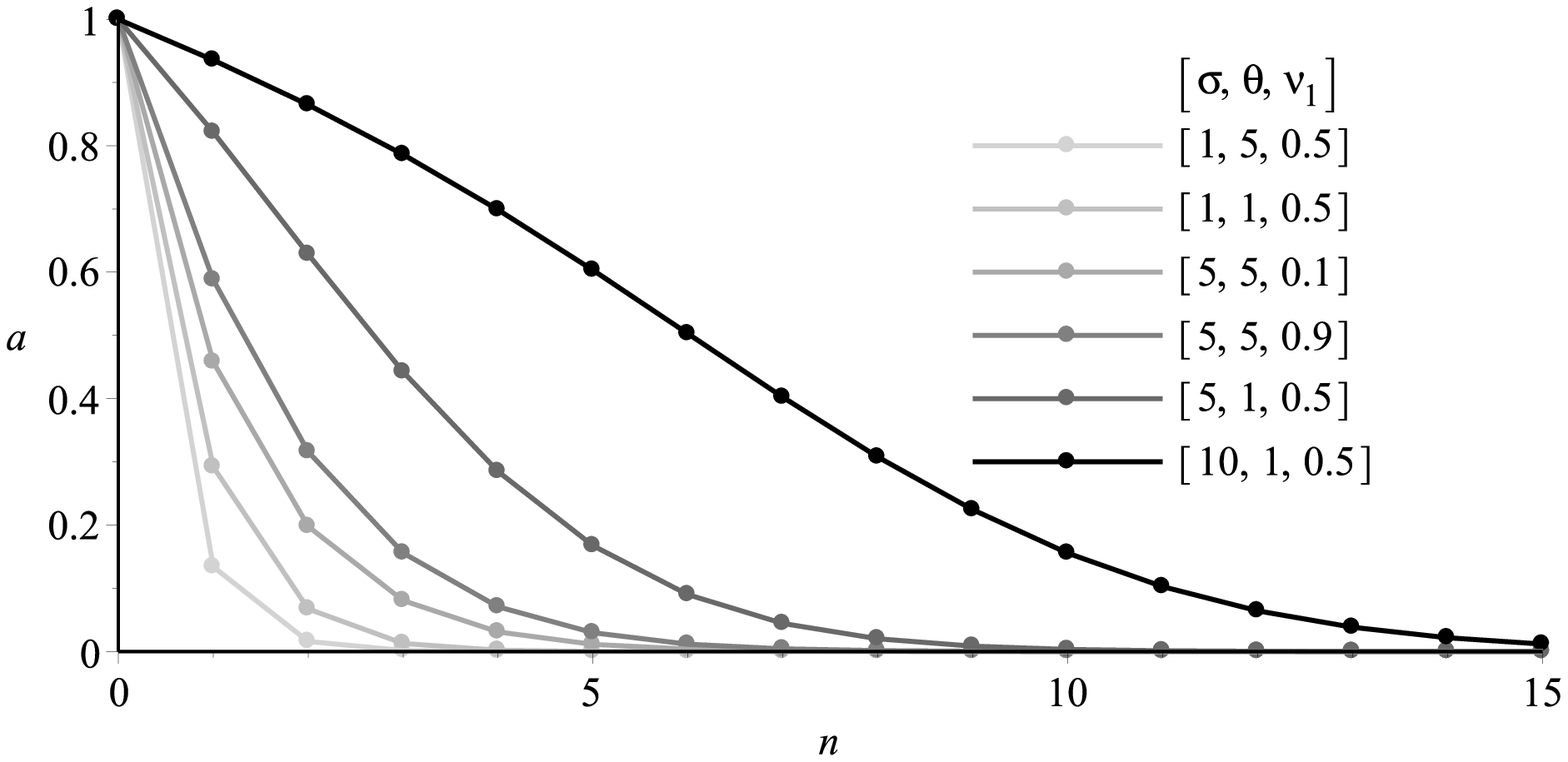}
\end{minipage}
\begin{minipage}[t]{0.8\textwidth}
\includegraphics[width=\textwidth]{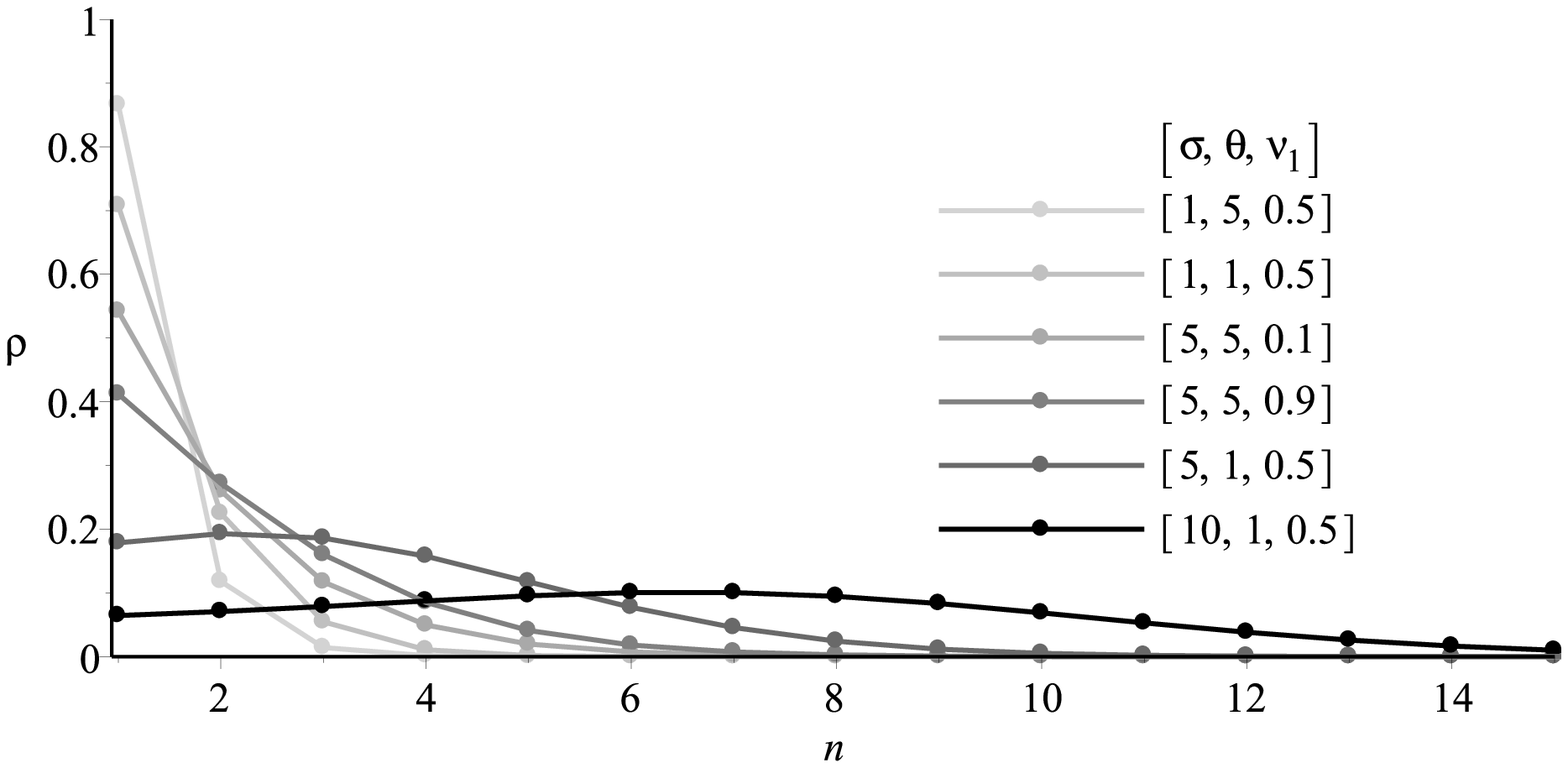}
\end{minipage}
\end{center}
\caption{Probability $h(x)$ that the immortal line at time $0$ carries the beneficial type, given the frequency of this type is $x$ (\textit{top}), tail probabilities $\alpha_{n}=\mathbb{P}(L_0 > n)$, $n \geq 0$, of the stationary distribution of the highest occupied level in the pruned LD-ASG (\textit{middle}), and  probability weights $\rho_n=\mathbb{P}(L_0=n)$ (\textit{bottom}), $n \geq 1$. Results are shown for different combinations of selection coefficient $\sigma$, mutation rate $\theta$, and mutation probability $\nu_1$ to type $1$.}
\label{bilderhxundrho}
\end{figure}

The (conditional) probability $h(x)$ that the immortal line at time $0$ carries the beneficial type does not only depend on the frequency $x$ of this type but also on three parameters: selection coefficient $\sigma$, mutation rate $\theta$, and mutation probability $\nu_1$ to the deleterious type. As shown in Fig.~\ref{bilderhxundrho}, some monotonicity properties apply. Since $h(x)=\sum_{n\geq 0} \alpha_n x(1-x)^n$ depends on the tail probabilities $(\alpha_{n})$ monotonically, an increase of $\alpha_{n}$ for all $n\in \mathbb{N}$ yields an increase of $h(x)$ as well. 
Let us now explain how the dependence of the tail probabilities on the three parameters can be understood in terms of the pruned equilibrium LD-ASG.\\
To this end, we consider the tail probabilities as functions of the parameters,
i.e., $\alpha_n = \alpha_n(\sigma, \theta,\nu_1)$. 

\begin{itemize}
	\item If $\sigma_1>\sigma_2$, then  
$\alpha_n(\sigma_1,\theta,\nu_1) > \alpha_n(\sigma_2,\theta,\nu_1)$. This is due to the fact that higher selection coefficients result in higher intensities of the Poisson point process $\Gamma^{\ast}$ of stars (compare Section~\ref{sec4_2}). Since each star indicates the birth of a line in the pruned LD-ASG, in distribution more lines are born, which increases the tail probabilities of the top level $L_0$.
	\item For $\theta_1>\theta_2$, one observes $\alpha_n(\sigma,\theta_1,\nu_1) < \alpha_n(\sigma,\theta_2,\nu_1)$. This is because each mutation results in deleting lines from the pruned LD-ASG (unless it is a deleterious mutation on the immune line or a beneficial mutation on the top line), and a higher mutation rate results
in more lines being cut away (in distribution). This decreases the tail probabilities for $L_0$.
	\item For $\nu_{1,1}>\nu_{1,2}$, one has $\alpha_n(\sigma,\theta,\nu_{1,1}) > \alpha_n(\sigma,\theta,\nu_{1,2})$. The reason is that increasing $\nu_1$
(at constant $\theta$) means replacing each
circle in a realisation of the Poisson point processes $\Gamma^{\circ}$ by a cross (with a given probability), which thus adds to $\Gamma^{\times}$. Since the pruning procedure can cut away more than one line at each circle but at most one line at each cross, we get, in distribution, more lines at higher $\nu_1$,
which explains the increased tail probabilities.
\end{itemize}
To summarise: For fixed $x$, the quantity $h(x)$, as a function of one of the parameters $\sigma$, $\theta$, and~$\nu_1$ (with the other two parameters being fixed), is strictly increasing in $\sigma$, strictly decreasing in~$\theta$ and strictly increasing in $\nu_1$. We will comment on the third of these monotonicity relations in Sec. \ref{monot}.

An illustration of the probability weights $(\rho_n)$ of $L_0$ (i.e., $\rho_n = a_{n-1} - a_{n}, n \geq 1$) for various parameter combinations is also included in Fig.~\ref{bilderhxundrho} (bottom).

\section{Taylor's representation of the CAT distribution via a boundary value problem}
\label{sec:bvp}
\begin{figure}[htdp]
\psfrag{0}{$0$}
\psfrag{1}{$1$}
\psfrag{r=0}{\hspace{-3.0ex} $r=0$}
\psfrag{b}{\hspace{-1.5ex} \small$X_{-s}=x$}
\psfrag{Xr}{$(X_r)$}
\psfrag{r}{$r$}
\psfrag{t}{$t$}
\psfrag{s}{$t=0$}
\psfrag{Y}{$(X_t)$}
\psfrag{T}{\small$T$}
\psfrag{a}{\hspace{-3.0ex} $t=-s$}
\begin{center}
\begin{minipage}[t]{0.45\textwidth}
\psfrag{X0=x}{\hspace{-0.5ex} \small$X_0=x$}
\includegraphics[width=0.8\textwidth]{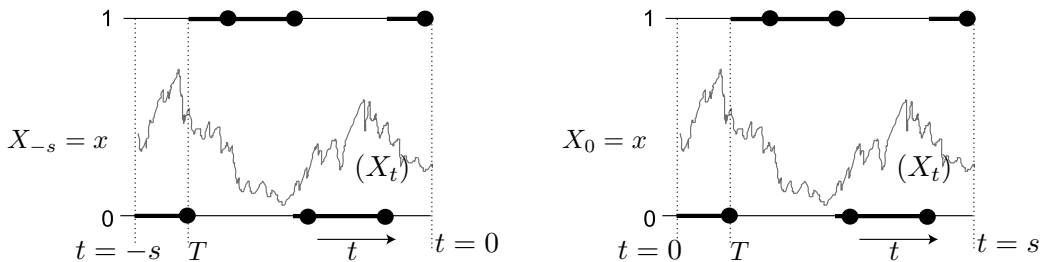}
\end{minipage}
\begin{minipage}[t]{0.45\textwidth}
\psfrag{X0=x}{\hspace{-4.0ex} \small$X_0=x$}
\psfrag{Y}{$(X_t)$}
\psfrag{t}{$t$}
\psfrag{a}{\hspace{-3.0ex} $t=0$}
\psfrag{b}{\hspace{-0.5ex} \small$X_0=x$}
\psfrag{T}{\small$T$}
\psfrag{s}{$t=s$}
\includegraphics[width=0.8\textwidth]{Taylor_2_new.eps}
\end{minipage}
\end{center}
\caption{Taylor's interpretation of \eqref{dir} after a time-reversal (between $t=-s$ and $t=0$) ({\em left}) and a time shift by $s$ ({\em right}), see Sec. \ref{hitprob}.  The type frequency path $X$ of the beneficial type figures as a background process.  In this realisation, the type of the common ancestor  is $0$ because the (backwards in time) jump at time $T$ is to type $0$. }
\label{Taylorpicture1}
\end{figure}

\citet{Taylor} shows by analytic methods (see also \citet{Kluth}) that the (conditional) common ancestor type probabilities $h(x)$ arise as the solution of the boundary value problem 
\begin{align}\label{dir}
\widetilde Ah(x) &= 0, \quad 0<x<1 \\
\lim_{x\to 0} h(x)&=0, \quad \lim_{x\to 1} h(x) = 1, \label{boundary}
\end{align}
where, for $\phi \in C^2([0,1], \mathbb R)$,
\begin{equation}\label{Atilde}
\widetilde A\phi(x) = A\phi(x) + \theta \nu_0 \frac {1-x}{x} \big (\phi(0)-\phi(x) \big ) +  \theta \nu_1 \frac {x}{1-x} \big (\phi(1)-\phi(x) \big ),
\end{equation}
and $A$ is the generator of the Wright-Fisher diffusion
\begin{equation}\label{Agenerator}
A\phi(x) := \left(x(1-x)\frac{d^2}{dx^2} + (\theta\nu_0(1-x)-\theta\nu_1x+\sigma x(1-x)) \frac{d}{dx} \right) \phi,
\end{equation}
see \citet[Proposition 2.4]{Taylor}. Together with his Proposition 2.5, Taylor then suggests the following interpretation of~\eqref{dir}: Given the frequency of the beneficial type at time $0$ is $x$, sample two lineages at time $0$, one of the beneficial type  and one of the unfavourable type, and trace them back into the past. He writes: ``By comparing this generator with that of the structured coalescent for a sample of size 2 ... [with two different alleles], it is evident that the type of the common ancestor has the same distribution as the type of the sampled lineage which is of the more ancient mutant origin.'' While Taylor here proposes to take the  type frequency path observed from time $0$ back into the past as a background process for the structured coalescent, his idea leads to a direct derivation and interpretation of~\eqref{dir} after a time reversal and a time shift (see Fig. \ref{Taylorpicture1}).  We take the chance to briefly explain this derivation here, as an add-on to \citet{Taylor} and to the approach developed in the previous sections. For this we start from the illustration in the right part of Fig. \ref {Taylorpicture1}.

\subsection{A representation of $h(x)$ as a hitting probability}\label{hitprob}
Let us fix $x \in \left(0,1\right)$ and consider the following two-stage experiment: In the first stage,  generate a random Wright-Fisher path $X=(X_t)_{t \geq 0}$  started in $X_0=x$ with generator~\eqref{Agenerator}. In the second stage, given $X$, we consider the ancestral lineage of an  individual  sampled at random from the population at a late time $s>0$. For $0\le u\le s$, let $J_{s-u}^s$ be the  type of that lineage at time $s-u$, i.e. $u$ units of time back from the time  of  sampling. In particular, $J_s^s \in \left\{0,1\right\}$ is the type at time $0$. We abbreviate  $b(x):= {(1-x)}/x$. Then, conditioned on the path $X=(X_t)_{t\ge 0}$ of the frequency of the beneficial type, the dynamics of $J^s := (J_u^s)_{0 \leq u \leq s}$ arises when restricting the structured coalescent investigated by \citet{BartonEtheridgeSturm} and \citet{Taylor} to a single ancestral lineage:
It is a $\left\{0,1\right\}$-valued jump chain with time-inhomogeneous backward-in-time jump rates $\lambda_{t}^{0,X}:=\theta\nu_0 b(X_{t})$ from $0$ to $1$ and $\lambda_{t}^{1,X}:=\theta\nu_1 / b(X_{t})$ from $1$ to $0$ at time $t=s-u \in \left[0,s\right]$, compare Fig.~\ref{Taylorpicture1}, right part. 
Conditioned on $X$, the processes~$J^s$, $s > 0$, can be coupled, i.e. constructed on the same probability space, by using two independent Poisson point processes $\Pi^{0,X}$ and $\Pi^{1,X}$ on $\mathbb R_+$ with  time-inhomogeneous intensities $\lambda^{0,X}$ and $\lambda^{1,X}$. This coupling works as follows: backwards in time, each of the processes $J^s$ jumps to $1$ at any point $\tau_0 < s$ of $\Pi^{0,X}$ (or remains in $1$ if it was already there), and jumps to $0$ at any point $\tau_1 < s$ of $\Pi^{1,X}$ (or remains in $1$ if it was already there).  Let us note that such a coupling would not be possible if one considers the time intervals $[-s,0]$ as in  Fig.~\ref{Taylorpicture1}, right part, since then the distribution of (an initial piece of) $X$ would vary with $s$. Thus, while the interpretation that goes along with the left part of  Fig.~\ref{Taylorpicture1} is more appealing from a biological point of view, the (mathematically equivalent) picture after the translation to the time interval $[0,\infty)$ (Fig.~\ref{Taylorpicture1}, right part) makes the analysis more convenient.

In the above-described coupling,  the common ancestor at time $0$ is of type $0$ if and only if $\lim_{s\to\infty} J^s_s =0$, which happens if and only if the point in the union of $\Pi^{0,X}$ and $\Pi^{1,X}$ that is closest to $0$ belongs to $\Pi^{1,X}$. As a matter of fact, such a closest point to $0$ exists: since $X_0  = x \in (0,1)$ and since the rates $\lambda^{0,X}$ and $\lambda^{1,X}$ are bounded as long as $X$ is bounded away from the points $\{0,1\}$, there is a minimal point $T_0$ in $ \Pi^{0,X}$ and a minimal point $T_1$ in $ \Pi^{1,X}$. Let $T:=\min\left\{T_0,T_1\right\}$. We have thus derived the representation
\begin{equation}\label{couplingfromthefuture}
h(x)=\mathbb{E}_x\left[\mathbb{P}\left(T=T_1\mid X\right)\right] . 
\end{equation}
%
%
\medskip
Now consider a jump-diffusion process $\widetilde X$ with generator $\widetilde A$ that starts in $x$, and let $\tilde T$ be the time of its first jump to the boundary. We then claim that 
\begin{equation}\label{initial}
 \Big((X_t)_{0\le t < T}, {\mathbb{I}}_{\{T=T_1\}}\Big) \stackrel d= \Big((\widetilde X_t)_{0\le t < \widetilde T}, \widetilde X_{\widetilde T}\Big),
 \end{equation}
where $\mathbb{I}$ is the indicator function.

To see this equality in law, recall that, given $X$,  points of $\Pi^{0,X}$ arrive at rate $\lambda^{0,X}$, while a  jump of $\widetilde X$ to the boundary point 0 occurs at rate $\lambda^{0,\widetilde X}$, and that, given $X$,  points of $\Pi^{0,X}$ arrive at rate $\lambda^{0,X}$, while a  jump of $\widetilde X$ to the boundary point 1 occurs at rate $\lambda^{1,\widetilde X}$.

 In view of \eqref{initial}, the representation \eqref{couplingfromthefuture} translates into
%
%
\begin{align}\label{theevent}
h(x) = \mathbb P_x( \widetilde X_{\widetilde T} = 1).
\end{align}
This shows that $h$ is a hitting probability of $\widetilde X$, and thus satisfies the Dirichlet problem \eqref{dir}. The boundary conditions~\eqref{boundary} are explained by the fact that the jump rates $\theta \nu_0 b(x)$ and $ \theta \nu_1 b(x)^{-1}$ converge to $\infty$ as $x$ converges to $0$ and $1$, respectively.

Thus the ``forward picture'' of Fig. \ref{Taylorpicture1} leads to the same characterisation of $h$ (in terms of a hitting probability of a jump-diffusion process) as Taylor's above-mentioned ``backward picture''. The reason for this is the  time-reversibility of the one-dimensional Wright-Fisher diffusion. This symmetry breaks down when the allele frequency dynamics are not invariant under time reversal (e.g., with multiple alleles and parent-dependent mutation). Still, a two-stage construction along the lines of Fig.\ref {Taylorpicture1}  might, in connection with a suitable ``coupling from the future'', lead to a related (but then more complicated) characterisation of the multitype analogue of $h(x)$.

\subsection{Discussion of  monotonicities  in the parameter $\nu_1$}\label{monot}
We have proved in Sec. \ref {monotonicities} that $h(x)$ is monotonically increasing in $\nu_1$ (for every fixed $x$). At first sight, this may seem paradoxical: how can it be that an increase in the mutation rate towards the disadvantageous type increases the probability that the common ancestor is of the beneficial type? The representation \eqref{couplingfromthefuture} resolves at least part of this paradox: 

 For fixed $X$, an increase of $\nu_1$ yields an increase of $\lambda^{1,X}$ and a decrease of $\lambda^{0,X}$. This results in an 
enhancement of $\mathbb{P}\left(T= T_1\mid X\right)$.  In other words, given $X$, the intensity of mutations ``back to the beneficial type'' increases as $\nu_1$ increases. 

Since, under $\mathbb P_x$,  $X_t$ (for $t>0$) has a tendency to become smaller as $\nu_1$ increases, the monotonicity of  \eqref{couplingfromthefuture} for fixed $X$ is not quite sufficient to prove the monotonicity of $h(x)$. Still, the explanation invoking the intensity of mutations ``back to the beneficial type'' gives some intuition why $h(x)$ should be increasing in $\nu_1$ - a result which we have derived in Sec. \ref {monotonicities} via the line counting process of the pruned equilibrium ASG.
 
Let us now turn  to the common ancestor type distribution. To this end, we make the dependence of the stationary density on $X$ (mentioned at the end of Sec. \ref {moranmodelsection}) explicit and denote it by $\pi_{\nu_1}(x)$, $0<x<1$.  Consider $g_{\nu_1} := \int_{\left[0,1\right]} h_{\nu_1}(x) \pi_{\nu_1}(x)dx$, that is the probability that, in the equilibrium of $X$, the common ancestor's type  is beneficial. We now have two opposing monotonicities: On the one hand, $h_{\nu_1}(x)$ increases with both $\nu_1$ and $x$; on the other hand, larger $x$ get lower weight under $\pi_{\nu_1}(x)dx$ as  $\nu_1$ increases. The monotonicity of $g_{\nu_1}$ is therefore not obvious. As noticed by Jay Taylor, it is plausible to conjecture 
that $g_{\nu_1}$ should be decreasing with $\nu_1$, which then would be an instance of Simpson's paradox.

\section{Conclusion}
The aim of this contribution was to find a transparent graphical method to identify the common ancestor in a model with selection and mutation, and in this way to obtain the type distribution on the immortal line at some initial time, given the type frequencies in the population at that time. 
This ancestral distribution is biased towards the favourable type. This
bias, which is  quantified in the series representation \eqref{hTaylor}, reflects its increased long-term offspring expectation (relative to
the neutral case). A closely related phenomenon is well known from multi-type branching
processes and deterministic mutation-selection models (see \citet{BaakeGeorgii} and
references therein).

Our construction relies on the following key ingredients. We start from the
\emph{equilibrium} ASG (without types), and from the insight  that the immortal
line is the one that is ancestral to the first bottleneck of this ASG.
Identifying this ancestral line had previously appeared to be difficult, since
it requires keeping track of the hierarchy of (incoming and continuing)
branches, which quickly may become
confusing. We overcome this problem by
\emph{ordering} the lines, in this way introducing a lookdown version of
the ASG and a neat arrangement of the lines according to their hierarchy.
Next, we {\em mark} the lines of the equilibrium ASG by the {\em mutation events} and, working backwards in time, apply a \emph{pruning} procedure, which cuts away those
branches  that  cannot be ancestral.
Finally, we \emph{assign types} at time $0$ to the lines of the
resulting pruned LD-ASG by drawing the types of its lines
from the initial frequency and thus
determine the type of the immortal line at time $0$.

This equilibrium
lookdown ASG is the principal (and new) tool in our analysis: backward in time, the top level
in the pruned ASG  performs a Markov chain whose equilibrium distribution can be
computed, and the tail probabilities of this equilibrium distribution are shown to obey the
Fearnhead-Taylor recursion. This  provides
the link to 
the simulation algorithm described by \citet{Fearnhead} for the common ancestor
type distribution in the stationary
case. More precisely, our Theorems~\ref{immortallineinlookdownASG}  and \ref{theorem_h}
together connect the LD-ASG to the simulation algorithm and thus provide a
\emph{probabilistic} derivation for it. At
the same time, they imply a  generalisation to an arbitrary rather
than a stationary initial type distribution. Furthermore, Theorem~\ref{theorem_h} sheds new light on the series representation for the conditional CAT
distribution, whose coefficients now emerge as the tail probabilities of
the number of lines in the pruned LD-ASG.
As a nice by-product, the graphical approach directly reveals
various monotonicity properties of the tail probabilities depending on the model parameters,
which translate into monotonicity properties of the common ancestor type distribution. 

We
believe that the pruned equilibrium (lookdown) ASG has potential for the graphical analysis
of type distributions and genealogies also beyond the applications considered in the present
paper. Let us also emphasise that, unlike  Fearnhead's original approach which builds
on the stationary type process (and unlike other pruning procedures that work in a
\emph{stationary typed} situation), we start out from the \emph{untyped} lookdown ASG which then is marked
and pruned, with the assignment of types at the fixed (initial) time being delayed until the
very last step of the construction. This is essential to be able to assign the types i.i.d. with
a given frequency, and in this way to arrive at the desired probabilistic derivation of the
conditional common ancestor type distribution.

\section*{Acknowledgements}
The authors thank Tom Kurtz and Peter Pfaffelhuber for stimulating and fruitful discussions. They are grateful to Martin M\"ohle, Jay Taylor and an anonymous referee for their help in improving the first version of this paper. This project received financial support from
Deutsche Forschungsgemeinschaft (Priority Programme SPP 1590 `Probabilistic
Structures in Evolution', grants no. BA 2469/5-1 and WA 967/4-1).

\section*{References}

\begin{RaggedRight}

\end{RaggedRight}

\end{document}